\documentclass[useAMS,usenatbib]{mn2e}
\usepackage {graphicx}

\newcommand{\lae}{\mathrel{<\kern-1.0em\lower0.9ex\hbox{$\sim$}}}
\newcommand{\gae}{\mathrel{>\kern-1.0em\lower0.9ex\hbox{$\sim$}}}

\title[Gravitational microlensing as a probe for dark matter clumps]{\bf
Gravitational microlensing as a probe for dark matter clumps}

\author[E. Fedorova, V.M. Sliusar, V.I. Zhdanov,    A.N. Alexandrov, A. Del Popolo, J. Surdej]{E. Fedorova$^{1, 2}$\thanks{ofedorova@ulg.ac.be}, V.M. Sliusar$^{2}$, V.I. Zhdanov$^{2}$\thanks{valeryzhdanov@gmail.com},   \and
 A.N. Alexandrov$^{2}$,  A. Del Popolo$^{3,4,5}$, J. Surdej$^{1}$.\\
\\$^{1}$ Institute of Astrophysics and Geophysics of Li\`ege University,
Quartier Agora
All\'ee du 6 ao\^ut 19C,
B-4000 Li\`ege,
 Belgium
\\$^{2}$ Astronomical Observatory of Taras Shevchenko National University of Kyiv, Observatorna 3,  Kiev, 04053, Ukraine
\\$^{3}$ Dipartimento di Fisica e Astronomia, University of Catania, Viale A.Doria 6, 95125, Catania, Italy
\\$^{4}$ INFN sezione di Catania, via S.Sofia, 95123 Catania, Italy
\\$^{5}$ International institute of physics, Universidade Federal do Rio Grande do Note, 59012-970, Natal, Brazil\\}

\def\a{a}

\begin{document}
\date{}
\pagerange{\pageref{firstpage}--\pageref{lastpage}} \pubyear{2014}
\maketitle \label{firstpage}

\begin{abstract}
Extended dark matter (DM) substructures may play the role of microlenses in the Milky Way and in extragalactic gravitational lens systems (GLSs). We compare microlensing effects caused by point masses (Schwarzschild lenses) and extended clumps of matter using a simple model for the lens mapping. A superposition of the point mass and the extended clump is also considered. For special choices of the parameters, this model may represent a cusped clump of cold DM, a cored clump of self-interacting dark matter (SIDM) or an ultra compact minihalo of DM surrounding a massive point-like object. We built the resulting micro-amplification curves for various parameters of one clump moving with respect to the source in order to estimate differences between the light curves caused by clumps and by point lenses. The results show that it may be difficult to distinguish between these models. However, some region of the clump parameters can be restricted by considering the high amplification events at the present level of photometric accuracy. Then we estimate the statistical properties of the amplification curves in extragalactic GLSs. For this purpose, an ensemble of amplification curves is generated yielding the autocorrelation functions (ACFs) of the curves for different choices of the system parameters. We find that there can be a significant difference between these ACFs if the clump size is comparable with typical Einstein radii; as a rule, the contribution of clumps makes the ACFs less steep.
\end{abstract}

\begin{keywords}
astrophysics, cosmology -- Gravitational microlensing, dark matter.
\end{keywords}

\section{Introduction}

Since the beginning of the last century, when some astronomers started to study the matter content of our neighborhood \citep{Opik,Kap,Jeans,OO,Zwicky}, many evidences have been collected leading one to believe that the Universe in which we live is mainly constituted by non-luminous  matter, whose existence is inferred through its gravitational effects on the remaining constituents of the Universe.
This ``missing matter", was dubbed ``dunkle materie" (dark matter), by \cite{Zwicky}. Nowadays, we know, according to Plank's data mission fitted with the $\Lambda$CDM model, that
the Universe is composed of 26.8\% non-baryonic dark matter (DM) and 68.3\% dark energy (represented by the $\Lambda$ term), and just 4.9\% ordinary matter \citep{Ade}. In this context, the
role of Gravitational Lensing (GL) can hardly be overestimated.  GL provides important evidences in favor of DM existence in galactic clusters \citep{Bullet, Clowe_2004,  Clowe_2006, Bradac_2006}. The other application of GL, following the idea by \cite{Paczyn}, deals with the searches for compact objects in the Galactic halo and inside the  Milky Way \citep{Alcock, Udalski,Aubourg}.
GL effects over a wide range of lens masses can give us the possibility of analyzing the DM substructure characteristics; especially it can give us a clue to solve the ``missing satellite'' problem \citep{Klypin, Moore, DLF}. Here we analyze the possibilities to study some properties of DM using mainly photometric information induced by gravitational microlensing  effects.

Following the terminology adopted by the GL community, at least three kinds of GL phenomena, characterized by different lens
masses and typical timescales, exist:
\begin{enumerate}
\item{macrolensing and the weak lensing by galaxies or groups of galaxies;}\label{type_i}
\item{mesolensing: lenses are globular clusters, dwarf galaxies or DM clusters with mass in the range 10$^3$ to 10$^9$ M$_{\odot}$ \citep{BB};}\label{type_ii}
\item{microlensing: lenses are stellar-mass objects.}\label{type_iii}
\end{enumerate}

As distinct from \ref{type_i} and \ref{type_ii} dealing with almost static situations or very slow processes, the
characteristic timescales of microlensing events in the Milky Way are of the order of weeks and in case of extragalactic microlensing
events, they are of the order of months or years. This makes image brightness variation effects due to \ref{type_iii} quite observable.

Based on the lens mass range/timescales, this terminology determines at the same time the possible observational appearance
of these effects, i.e. multiple correlated static images with signal arrival time delays in the macrolensing case,
slowly-varying distorted images due to mesolensing, and high amplification events on lightcurves when the source crosses the
caustic of a microlens. Dark matter can manifest itself at the three  mentioned levels, namely:
\begin{enumerate}
\item{large haloes of DM with mass greater than $10^9 M_{\odot}$ can play the role of macrolenses; the most prominent example is the "Bullet
Cluster"\citep{Bullet};}
\item{DM subhaloes of intermediate masses can play the role of mesolenses, causing the anomalous flux ratios,
image distortions \citep{C} and additional time delays \citep{KM} in extragalactic gravitational lens systems. Several candidates to
show anomalous flux ratios are known today, e.g., B2045+265 \citep{MK}, RX J1131--1231, B1608+656, WFI 2033 -- 4723
\citep{CKN}, B1938+666 \citep{ML}, and we would especially like to stress here the famous GLS Q2237+0305 for its sharp high-amplification
microlensing events \citep{MMM}.}
\item{Clumps of DM with stellar masses can play the role of microlenses, leading to both photometric (high amplification events)
and astrometric (jump-like shifts of brightness centroid of the images of the microlensed source) appearances. The main difference
between DM microlensing and the "usual" one due to stars or black holes lies in the non-negligible size of the DM clumps (non-DM
microlenses are always considered as point masses).  }
\end{enumerate}

In this paper, we pay attention  to the last item, i.e. microlensing by extended clumps of stellar masses. We note that
continuous observations by EROS, OGLE and other groups \citep{EROS,OGLE} neither revealed any sign of these extended
clumps in the Galactic halo, nor provided any proof of such structures inside the Galaxy. The overwhelming majority of the light
curves observed for Galactic microlensing events are well described by the gravitational field of stars (and, sometimes, of planetary
mass objects). Therefore, it seems that there is no room for the stellar mass extended clumps. However, in this paper we show that
the light curves due to extended clumps of stellar mass objects can mimic the light curves caused by ordinary compact objects. Therefore,
more detailed investigation is needed in order to rule out (or to confirm) the existence of these DM clumps that may either not
be very numerous (so that we do not have these clumps within our Solar system) or not so dense to be observable.

Photometric signatures of DM substructure via gravitational microlensing have been widely discussed; see \cite{MJOW,MK,O} and
references therein. There are also some investigations of variability of spectral line profiles \citep{MMM}.
\cite{PaczWambs} derived amplification distributions for static gravitational macro-lensing with a non-constant surface mass density, which
includes the cases of a stochastic system of Gaussian clumps and clumps in the form of truncated singular isothermal spheres.
\cite{ZS}, and \cite{Z} have considered a model of microlensing due to a non-compact neutralino star.

In this paper we use several other models of DM clumps to analyze the observational appearance of DM microlensing within different cosmological models. We focus our attention on item \ref{type_iii} to compare the observational appearances of microlensed sources due to point-like and finite-size DM clump deflectors and determine how one can distinguish an extended DM
clump microlensing event from a ``regular'' one in case of Galactic and extragalactic structures.  The main question we want to answer concerns the possibilities to detect
signals from these  putative extended mass structures and to estimate the accuracy needed to characterize such effects mainly from observed light curves.

In Section 2, we briefly describe the existing DM models. In Section 3, we propose a "toy model", which describes, for special
choices of the parameters,  microlensing by various types of extended structures. The model is based on  the deflection angle
$\mathbf \alpha \sim {\mathbf r}(r^2+r_s^2)^{-\a/2}$, where $r_s$ characterizes the effective size of a clump core\footnote{The
corresponding surface mass density is $\rho \sim (1-a/2)(r^2+r_s^2)^{-a/2}+(a/2)r_s^2(r^2+r_s^2)^{-(1+a/2)}$}. For
different choices of parameters, this model reproduces the effects either due to point microlenses, or  CDM minihaloes with a cusped
density profile, or SIDM clumps with a cored density profile. In Section 4 we discuss the superpositions of extended and point-mass
deflecting objects. In the case of isolated microlenses these models can be used to describe  microlensing events within the Galaxy.

We perform a numerical comparison of these events  considering different microlens models. Namely, we calculate the "amplification curves",
that is the dependence of the total amplification of a microlensing system versus time, as the lens moves with respect to the
line-of-sight to the source. We also pay some attention to  the trajectories of the brightness centroid of the microlensed images,
because the fast progress in the positional accuracy measurements\footnote{e.g., projects of radio interferometry
in space. Note that astrometric accuracy is typically several times larger than the resolution; i.e. one can  achieve a
microarcsecond level in positioning, though different  images of the microlensed source cannot be resolved at this level.} gives us
a hope to detect the astrometric signatures of  gravitational microlensing  in the near future.

In Section 5 we consider the observational behavior of clumps in extragalactic GLSs. Namely, we study the statistical effects of
gravitational microlensing due to a stochastic system composed of extended DM clumps  and point masses. There is a number of papers dealing with statistical
subjects of microlensing systems; see, e.g., \cite{SchnEhlFal,Seitz1994, Neindorf, Dobler2007, Schmidt2010}. Our technique is essentially the same as that
used by \cite{PaczWambs} and \cite{WambPaKa}, followed by a number of authors; cf. especially considerations by \cite{Metcalf,Chiba,Dal-Koch,ScW} dealing with brightness variations in connection to the problem of anomalous brightness ratios. Unlike in the previous works we deal with the autocorrelation functions (ACFs) of the amplification curves for our concrete model of a microlensing system.  We generate an ensemble of amplification curves for a fixed input of randomly distributed clumps and point mass microlenses in the total optical depth; this
enables us to derive ACFs for these amplification curves as a function of different clump contribution, size and  in presence of an
external shear. Finally, in Section 6 we discuss the results.

\section{Properties of various DM models}

DM microstructure still remains the subject of debates; there are many hypotheses about DM particles \citep{SW,ZR}. Cold dark matter
is supposed to contain heavy particles, weakly interacting with each other and with baryonic matter (axions or WIMPs), and warm
dark matter (WDM) consists of light, fast-moving particles (sterile neutrinos, CHAMPs, neutralinos or gravitinos), which tend
less to form small-scale compact structures \citep{SSMM, DP}. Self-interacting dark matter can be considered as a particular kind
of cold one with a nonzero impact distance for the interaction between DM particles \citep{K,R}.

Recent cosmological N-body simulations within CDM \citep{SKM1,DMS, D2, Springel, Stadel, Vog} and WDM models \citep{KAP, SSMM}
have shown that dark matter is not distributed in space homogeneously, instead it forms more or less compact structures
surrounded by less dense continuously distributed DM. The most massive structures have been recognized for quite a long time
already: these are galactic and galaxy cluster DM haloes. But numerical simulations demonstrate that these massive structures
are also inhomogeneous: they contain smaller compact substructures over a wide range of masses. Such massive haloes containing a
hierarchical substructure of dark matter are called host haloes. These haloes also contain smaller substructures, etc. However, one
should note that only the existence of the more massive members of this hierarchy of DM structures has been
proven from the results of observations. The lower estimated limit on the substructure masses depends on the particular kind of
considered hypothetical DM particles, varying over a very wide range from $10^{-12}M_{\odot}$ for CDM up to $10^8M_{\odot}$ for
WDM \citep{BDE, BDE2, D2, Springel}. To clear out the situation and the terminology used to characterize masses and scales of  DM
substructures, we have summarized here in Table \ref{table0} the used nomenclature.

\begin{table}
 \caption{Dark matter hierarchy within the CDM/SIDM cosmologies.}
\begin{tabular}{|p{50pt}| p{80pt}|p{50pt}|} \hline
Term & Objects/scales  &   Mass range\\\hline Superhalo & Galaxy
clusters, 100 Mpc & 10$^{13}$-10$^{16} M_\odot$ \\\hline Halo &
Galaxies, Mpc & 10$^{9}$-10$^{11} M_\odot$ \\\hline Subhalo &
Dwarf galaxies or satellites, kpc & 10$^{4}$-10$^{9}
M_\odot$\\\hline Minihalo or clump & Stars & 10$^{-3}$-10M$_\odot$
\\\hline Microhalo & Planets &  10$^{-8}$-10$^{-4}
M_\odot$\\\hline Nanohalo or primordial halo & &
10$^{-18}$-10$^{-9}M_\odot$\\\hline
\end{tabular}
\vspace{0.05in} \label{table0}
\end{table}

DM substructure depends very significantly on the cosmological model. Within CDM models, the substructure formation process can
be described as ``bottom-up'': the smallest DM clumps were formed first (therefore the nanohaloes with masses $<10^{-11}M_{\odot}$
are often referred as primordial) and do not contain any finer substructure \citep{CLB}. Then, in the merging process more and
more massive structures up to galactic and cluster scales (10$^{10}$-10$^{15}$ M$_{\odot})$ were gradually formed. Due to
various damping scales the lower limits on substructure mass can vary over a wide range: from 10$^{ - 4}$-10$^{- 12}$ M$_{\odot}$
for various kinds of WIMPs (including neutralinos), down to 10$^{- 18}$-10$^{ - 20}$ M$_{\odot}$ for axions. The mass density
distribution in DM structures within the CDM model is often assumed to follow  the cusped NFW profile of \cite{NFW}.

The best accepted alternative to the CDM cosmology is the velocity-dependent SIDM (self-interacting-dark-matter). This model
predicts identical behavior of the DM substructure over various scales and identical mass fractions, thus, the only difference
lies in the density profile of a DM clump. The SIDM clump has a core and its density follows a Burkert profile \citep{R}, or
pseudo-isothermal sphere \citep{K}, contrarily to the cuspy coreless NFW profile typical for CDM clumps.

The set of warm dark matter models differs completely from the two mentioned above when we keep in mind gravitational microlensing processes.
The substructure formation in WDM models is a complex hybrid; its leading role is played by the ``top-down'' formation process \citep{KAP},
i.e. bigger haloes and filaments were formed first, and smaller structures formed later, as a
result of fragmentation processes in WDM filaments \citep{KDGS}. The lower limit on the subhalo mass is significantly higher than in CDM/SIDM ones: e.g., $10^6- 10^{8}$ $M_{\odot}$ for gravitinos
and even $10^{11}  M _{\odot}$ for axinos \citep{HIT}. Thus in the WDM cosmology, we may predict the non existence of DM gravitational microlensing. Even despite that
it was recently shown \citep{PRP} that the WDM structure formation rather follows the hybrid scenario than the ``top-down'' one, the effects of fragmentation and collapse
play a significant role only in massive haloes formation, and thus the situation with dark matter induced gravitational microlensing in WDM cosmology is not significantly altered.

The DM subhalo mass function (SHMF) $n(m,M_{0})$, such that $n(m,M_{0})dm$ represents the space number of
substructures with a mass in the range ${\{m,m+dm\}}$ in the host DM halo with a mass $M_{0}$ is strongly sensitive to the DM context (i.e. warm, cold, collisionless, repulsive and so on).
The SHMF appears to provide the essential source of crucial information both in cosmology and elementary particle physics. The numerical
simulations like ``Millenium'' \citep{SXXXX} or ``Via Lactea'' \citep{CLB} allow us to determine it only at higher subhalo masses
(mainly $>10^{6} M_{\odot})$ than typical microlens masses (i.e. not greater than $100M_{\odot})$. However for lower masses of the
substructure we can use here the extrapolated mass function obtained by \cite{LAK}:
\[
f(m) = 0.1\frac{{\log} \left( {M_{max}/m} \right)}{{\log} \left(
{M_{max}/M_{ch}} \right)}
\]
\noindent where $M_{\max } = 0.01M_{\odot}$  and $M_{ch} = 10^7M_\odot $. Thus, the total mass in substructures within the CDM model (with a lower limit of $10^{ - 6}$ $M_{\odot})$ is 52{\%},
following  \cite{LAK}. This value is in good agreement with the 50{\%} of dark matter in the Galactic halo in substructures
obtained from the hydrodynamical simulations by  \cite{DMS}. Using this formula, one can easily find that for the lower limit of
$10^{ - 3} M_{\odot}$ (which can be considered as a reasonable limit for microlens mass) it is 43{\%}, and for $10M_{\odot}$ (let
us consider this value as an upper limit on microlens mass) it is 30{\%}. Thus within the desirable range $(10^{ - 3} -
10)\,M_{\odot}$ of the substructures we have 13{\%} of the total DM mass (i.e. $\sim  1.3\cdot 10^{11}M_{\odot}$ is in compact
stellar-mass subhaloes). If we take into account the fact that the larger subhaloes (with mass greater than $~10M_{\odot}$) also
contain finer substructures, this value can be even larger.

However, estimating the probability of the DM subhalo microlensing we should take into account that up to 90{\%} of the primordial
subhaloes had to be fully or partially disrupted by tidal forces of stars, thus the percentage of the clumped matter in the areas
containing stars cannot exceed 10{\%} \citep{SKM2, SWOB}. Thus we can expect to trace the DM-induced high amplification microlensing
events in the Galactic DM halo (as well as in extragalactic systems) rather than in its luminous parts. The probability of the
Galaxy stars being microlensed by a DM microhalo is much lower then. That is why we can hardly expect to find compact DM clumps
in the vicinity of the Solar system.

However, at the same time, in the luminous parts of a galaxy we can face the situation when a star (or a star-mass black hole) is surrounded by a dense cloud,
formed by dark matter of a former clump, disrupted by star/black hole tidal force and trapped by its gravity. If such a clump is dense enough and thus has an optical depth large enough (overcritical
convergence), we can observe the UCMH \citep{Zhang, BDE2}, and one of the possible results of gravitational microlensing by such an object looks like an eclipse of the source image. The role of optical
depth/convergence in microlensing processes was analyzed by \cite{P}. In the simplest case such an object can be modeled up by a Schwarzschild or Chang-Refsdal lens with overcritical convergence.

\section{ Extended clump microlensing in the Galaxy}\label{extended_clump_models}

\subsection{The lens equation}

Speaking about the Milky Way systems, we confine ourselves to the simplest cases of circular symmetric mass distributions though
investigations of double stars or planetary systems are also important. One can get some insight into the latter case from
considerations of clump models with an external shear \citep{KFNT}. In this Section, we use the lens equation in its
normalized form; the distances will be expressed in units of a typical length scale $R_*$, where for the case of the point mass
$M$ or cored clump microlens $R_*^{2}=4G M D_{ds}/(c^2 D_d D_s)$ (in case of a point mass this is the radius of the Einstein ring)
and for the cusped clump $R_*^\a=8\pi G \rho_0  D_{ds}/[c^2 D_d D_s (2-\a)]$; here $\,D_s$ is the distance between the source and
the observer,  $\,D_d$ the distance between the deflector and the observer, and  $D_{ds}$ the distance between the source and the deflector. Furthermore $\mathbf y$ is the normalized angular source position, $\mathbf r=\{x_1,x_2\}$ is the normalized angular image position, $r=|\mathbf r|$. We remind here that (after normalization) both these vectors
are defined in the lens plane.

The normalized lens equation is:
\begin{equation}
\label{eq0z} \mathbf{y}=(1-\sigma)\,\mathbf{r}-\mathbf\alpha(\mathbf r),
\end{equation}
\noindent where the deflection angle
\begin{equation}
\label{phi} \mathbf \alpha(\mathbf r) = \frac{\mathbf r}{r^2} \int_0^r d{r'} \,r'\rho(r')
\end{equation}
is supposed to be normalized to a corresponding length scale, $\sigma=const$ stands for a convergence (optical depth) that can be due to, e.g., a background DM; $\mathbf y=\{y_1,y_2\}$.

Let us write $ \mathbf \alpha(r)=\mathbf{r}/r^\a$, where for the cusped clump we assume $\a<2$ to provide the convergence in the
integral of Eq.(\ref{phi}), and for the point microlens $\a=2$. For the cored clump we use a generalized expression $\mathbf
\alpha(\mathbf r)=\mathbf r/(r^2 + r_s^2)^{\a/2}$ that formally covers the previous cases if we put $r_s=0$. Therefore, we consider the lens equation

\begin{equation}
\label{eq1z} \mathbf {y } = \mathbf {r}\left(1-\sigma-  R^{-\a}\right), \quad R=\sqrt{r^2+r_s^2},\quad a>0. \end{equation}

The determinant of the lens mapping (\ref{eq1z}) is $D=\det \{\partial y_i/\partial x_j\}$. It  factorizes as follows:
\begin{equation}
\label{det}
D(r)=\varphi_1(r)\varphi_2(r),
\end{equation}
where
\begin{equation}
\label{factors}
\varphi_1(r)=1-\sigma -\frac{1}{R^{\a}}, \quad
\varphi_2(r)=1-\sigma +
\frac{\a-1}{R^{\a}}-\frac{\a r_s^2}{R^{\a+2}}.
\end{equation}
The image parity is defined by the sign of $D$.

The relative amplification\footnote{this is normalized to the amplification $(1-\sigma )^2$
when the source is far from the lens.} of the image at $\mathbf r$ is $\mu(\mathbf r)=(1-\sigma )^2/|D(r)|$.

\noindent Taking the absolute value of both sides of Eq. (\ref{eq1z}) yields
\begin{equation}\label{eq1mod}  y  = |f(r)|\,, \end{equation}
where $f(r)\equiv r\,[ 1-\sigma-  (r^2+r_s^2)^{-\a/2}], \quad
y=|\mathbf y|$. Evidently, if $r$ is a solution of
Eq.(\ref{eq1mod}), then the solution of Eq.(\ref{eq1z}) is either
$\mathbf r=\mathbf n_{\mathbf y} r$ or $\mathbf r=-\mathbf
n_{\mathbf y} r$, where  $\mathbf n_{\mathbf y}=\mathbf y/y$.

Our first goal will be to estimate the number of solutions of Eq.(\ref{eq1z}), i.e. the number of microlensed images for different source positions,
and to find the caustics of the lens mapping (\ref{eq1z}) where the lensed image of a point-source gets infinitely amplified. The problem is reduced to investigate  the monotonicity of the r.h.s. of Eq. (\ref{eq1z}) and the zeros of the determinant $D(r)$
in Eq. (\ref{det}).
Note that any root of $\varphi_2(r)$, if it exists, is smaller than the root of $\varphi_1(r)$ (for $\a>0$) because of the monotonicity of $\varphi_1(r)$ and due to $\varphi_2(r)=\varphi_1(r)+\a r^2/R^{\a+2}$.

Simple calculations yield
\begin{equation}
\label{dydr}  \frac{df(r)}{dr} = \varphi_2(r), \quad
 \frac{d\varphi_2(r) }{dr} =-\frac{\a r}{R^{\a+4}}\left[(\a-1)r^2-3r_s^2\right],
 \end{equation}
whence for $\a>1,\,r_s>0$ we infer the existence of the only inflection point $r_{infl}=r_s\sqrt{3/(\a-1)}$ of the curve $y= f(r)$ and there is no inflection for $\a<1$.

\begin{figure}
\includegraphics[width=3.5in,height=2.7in]{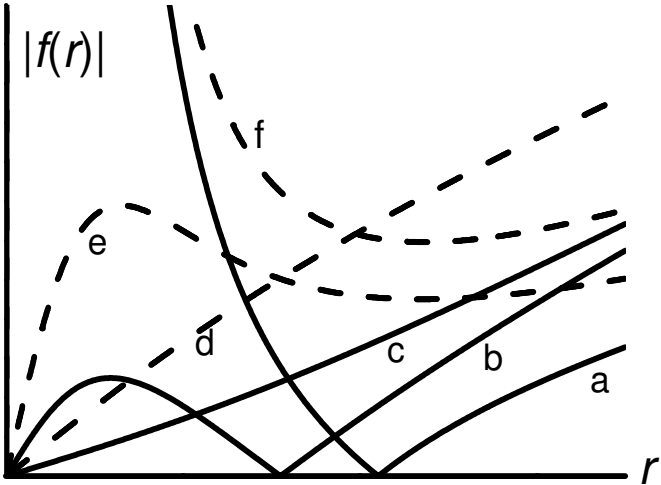}
\caption{Qualitative behavior of the r.h.s. of Eq. (\ref{eq1mod}) for the subcritical case ($0\le\sigma <1$:  solid; a,b,c) and the overcritical
case ($\sigma >1$:  dashed; d,e,f).  Here we show the subcritical cases (a) $\a>1,\,r_s=0$, (b)   $0<r_s<r_{s,cr}$,  (c) $r_s>r_{s,cr}$;
in case of (b) and (c) there may be an inflection point  that does not affect the number of  images. The
overcritical cases: (d) $0< \a<1$;  (e) $\a>1,\, r_s<r_{s,cr1}$; (f) $r_s=0, \, \a>1$.}\label{sideb}
\end{figure}

\subsection{Cored clump, subcritical $\sigma $}

Let us first consider the case   $r_s\ne 0$ and the subcritical values of  $\sigma $: $0\le\sigma <1$. Further we denote $r_{s,cr}=(1-\sigma )^{-1/\a}$.

 For the case  $r_s<r_{s,cr}$, the first factor $\varphi_1(r)$ of the r.h.s. in Eq. (\ref{det}) equals   zero for $r=r_{z}\equiv [(1-\sigma )^{-2/\a}-r_s^2]^{1/2}$.
 This corresponds to a circular critical curve with a radius $r_{z}$ around the origin of the image plane, which is mapped onto an isolated caustic point $\mathbf y=0$ in the source plane.
 Thus, there is a solution for Eq. (\ref{eq1z}) with $\mathbf y=0$.
For $r_s > r_{s,cr}$ the function $\varphi_1(r)$ is always positive.

For the second factor $\varphi_2(r)$ in Eq. (\ref{det}) we also have the condition\footnote{This condition is formally the same as the one for
the existence  of a root of $\varphi_1(r)$} $1-\sigma <r_s^{-\a}$ (i.e. $r_s<r_{s,cr}$) for the existence of $r_c<r_z$ such that $\varphi_2(r_c)=0$. Indeed, under this
condition, $\varphi_2(0)=1-\sigma -r_s^{-\a}<0$ and $\varphi_2(r_z) >0$ have different signs.
This is a necessary and sufficient condition for a unique value of $r_c$ to exist on $(0,r_z)$.
The proof of the uniqueness is somewhat different for $0<\a<1$ and $\a>1$. For $0<\a<1$ the function $\varphi_2(r)$ is monotonically increasing.
For $\a>1$ one should use Eq. (\ref{dydr}) and take into account the existence of the unique inflection point for
$f$: we observe that  $\varphi_2(r)$ is monotonically increasing for $r<r_{infl}$ (therefore there can be only one zero in this region provided that $r_s<r_{s,cr}$);
after passing through a maximum at $r=r_{infl}$  it decreases and it is positive:  $\varphi_2(r)>\varphi_2(\infty)=1-\sigma >0$.

It is easy to see that $r_c<r_z$, when it exists, is the radius of a circular critical curve that is mapped onto a circular caustic with radius $y_c=|f(r_c)|$. There exist two additional
images of a point source for the case $y<y_c$. The problem can be easily treated using  the graph of $|f(r)|$ (Fig. \ref{sideb}, b,c) and
taking into account the fact that $d|f|/dr$ is either equal  to $-\varphi_2$ or to $\varphi_2$. For $r_s>r_{s,cr}$ the function $|f(r)|$ starting at $r=0$ is monotonically
increasing, therefore, for any $\mathbf y$ there is a unique solution $r(y)$ of Eq. (\ref{eq1mod}).
The (unique) solution of Eq. (\ref{eq1z}) is  $\mathbf r = \mathbf n_{\mathbf y} r(y)$ having positive parity.

For $r_s<r_{s,cr}$ we have $f(r)<0,\,\, r\in (0,r_z)$. The function $|f(r)|$ has  only one maximum $y_c=|f(r_c)|$. For $r>r_z$ the r.h.s. of (Eq. \ref{eq1mod}) is a monotonically
increasing function. Therefore, for $0<y<y_c$ there are three solutions for Eq. (\ref{eq1mod}): two solutions $r=r_i\in (0,r_z), i=1,2;\,r_1<r_2$, which yield two solutions of the vectorial lens
equation (\ref{eq1z}) $\mathbf r=-\mathbf n_{\mathbf y} r_i$ ($r_1$ with positive parity, $r_2$ with negative parity); and one solution $r_3>r_z$ yielding  the solution $\mathbf r=\mathbf n_{\mathbf y} r_3$ with positive parity for Eq.  (\ref{eq1z}).

To sum up, for $0\le\sigma <1$ we have a unique image  for the case  $r_s>r_{s,cr}$; in addition, for $y=0$ there is an image\footnote{for  completeness, we note that
the trivial solution $\mathbf{r}=0$ for $\mathbf{y}=0$ exists, if $r_s\ne 0, \, \forall a>0$, and if $r_s= 0, \,  a<1$. } at $r=0$. All images have a finite brightness.
In the opposite case, for $0<r_s<r_{s,cr}$ there is a circular caustic with radius $y=y_c$ and there are three images, if the source is inside this caustic $0<y<y_c$,
and one
image, if  $y>y_c$. Two images acquire an infinite amplification when $y\to y_c-0$ and then disappear after crossing $y_c$. In addition, there is a caustic point $y=0$
corresponding to
a ring image like the Einstein ring emerging in case of a point mass lens.

\begin{figure}
\centerline{\includegraphics[width=3.0in,height=2.5in]{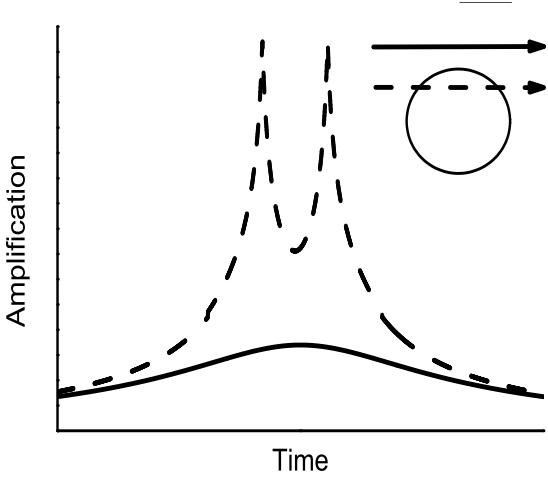}}
\caption{Qualitative behavior of the amplification curves, the subcritical case, $0<r_s<r_{cr}$.
The upper curve (dashed): the point source moves along the straight line with the impact parameter $y<y_c$
yielding two caustic crossings. The lower curve (solid): larger $y>y_c$, no caustic crossings.  In the right upper corner: the trajectories with respect to the circular caustic of the lens.}\label{Two_lightcurves}
\end{figure}

The caustics with source tracks and corresponding amplification curves are shown in Fig.\ref{Two_lightcurves}.

\subsection{Cored clump, overcritical $\sigma $}
For $\sigma >1$ ($r_s\ne 0$), let us first consider the case $0<\a<1$. In this case in equation (\ref{eq1mod}) $|f(r)|\equiv r[\sigma -1+ (r^2+r_s^2)^{-\a/2}]$ is a
monotonically increasing function (see Fig. \ref{sideb}, d, dashed curve). Its derivative is $  d|f(r)|/dr =- \varphi_2(r)>\sigma -1>0$. The lens mapping has no
caustics and critical curves; for any $\mathbf y$ there is always a unique solution $r$ for Eq.(\ref{eq1mod}) and a
unique solution  $\mathbf r =- \mathbf n_{\mathbf y} r$ (positive parity) for the vectorial lens equation (\ref{eq1z}).

For $\sigma >1$, $\a>1$ there is an inflection point of the function $|f(r)|\equiv -f(r)$ according to Eq. (\ref{dydr}). Taking into account the sign of this inflection, we see that there exists a minimum for $ d|f(r)|/dr$ at $r=r_{infl}$, which is equal to
\[f'_{min}=\sigma -1-\frac{2}{r_s^{\a}}\left(\frac{\a-1}
{\a+2}\right)^{\a/2+1}.
\]
Then new roots of $ d|f(r)|/dr$ appear when
\[ r_s<r_{s,\,cr1}=\left(\frac{2}{\sigma -1}\right)^{1/\a}\left(\frac{\a-1}
{\a+2}\right)^{(\a+2)/(2\a)}.
\]
If $r_s\ge r_{s,\,cr1}$, the function $|f(r)|$ is monotonically increasing (not shown in Fig. \ref{sideb}).
Let these new roots be $r_{c,1}$ and $r_{c,\,2}$\, with $r_{c,1}<r_{c,\,2}$, $r_{c,1}$ being a point of a local maximum of $|f(r)|$
and $r_{c,\,2}$ being a point of a local minimum (see Fig.
\ref{sideb}, e, dashed). The roots correspond to the radii of the  critical curves, and $|f(r_{c,1})|, |f(r_{c,\,2})|$
are the radii of the circular caustics. For $|f(r_{c,1})|<y<|f(r_{c,2})|)$ Eq.(\ref{eq1mod}) has three solutions $r_1<r_2<r_3$
correspondingly with a positive, negative and positive parity yielding three lensed images of a point source, and there is only one image with positive parity
otherwise. The solution $r_1(y)$ is
extended for  $y<|f(r_{c,1})|$ and the corresponding image has an infinite amplification when $y\to|f(r_{c,\,1})|-0$; in
this case $r_1(y)$ and $r_2(y)$ tend to $r_{c,\,1}$   and then disappear after $y$ crosses the value $|f(r_{c,\,1})|$.
The solution $r_3(y)$ is extended to all values of $y>|f(r_{c,\,2})|$; it has analogous properties near the other
caustic for $y\to|f(r_{c,\,2})|+0$. In this case $r_2(y)$ and $r_3(y)$ tend to $r_{c,\,2}$  and then disappear as $y$
decreases. We have for the solutions of Eq.(\ref{eq1z}), $\mathbf r=-\mathbf n_{\mathbf y} r_{\,i},\,\, i=1,2,3$.

\subsection{Cusped clump: $r_s=0$}
Here we have $f(r)=r(1-\sigma -r^{-\a})$. For $0<\sigma <1$,  the function  $f(r)<0$ for $r<r_{z1}=(1-\sigma )^{-1/\a}$; $r_{z1}$ is the radius of a critical curve
corresponding to the caustic point $y=0$.

For $0<\sigma <1$ and $0<\a<1$, the modulus $|f(r)|$ increases for $r<r_{max}=[(1-\a)/(1-\sigma )]^{1/\a}$ and then decreases to zero for $r\to r_{z1}$.
The value $y_c=|f(r_{max})|=\a \left[(1-\a )/(1-\sigma)\right]^{1/\a-1}$ is the radius of a circular caustic. The behavior of the graph of this
function is roughly the same as that illustrated by the  solid curve  in Fig. \ref{sideb}, b. There are  three solutions
for Eq. (\ref{eq1mod}) with  $0<y<y_c$: namely, two solutions $r_i, \,i=1,2,\, r_1<r_2$ for Eq. (\ref{eq1mod}),
$r_i<r_{z1}$ corresponding to images at $\mathbf r=-\mathbf n_{\mathbf y} r_i$ (with positive ($r_1$) and negative ($r_2$) parity, and one solution $r_3>r_{z1}$
corresponding to image at $\mathbf r=\mathbf n_{\mathbf y} r_3$  with positive parity. The latter solution remains valid for $y>y_c$.

For $0<\sigma <1$, $\a>1$, the function $f(r)$ is negative for $r\in(0,r_{z1})$, the modulus $|f(r)|$ decreases along this interval and  it increases for $r>r_{z1}$.
The behavior of $|f(r)|$ is represented by the curve (a) in Fig. \ref{sideb}. In this case there are always two solutions of Eq. (\ref{eq1mod}) for $r_1<r_2$
corresponding to images at points $\mathbf r=-\mathbf n_{\mathbf y} r_1$, $\mathbf r=\mathbf n_{\mathbf y} r_2$.

For $\sigma >1$ and $0<\a<1$ the function $|f(r)|$ increases for all values of $r>0$ starting from zero (like the dashed curve (d) in Fig. \ref{sideb}).
There is only one solution and one image for all $y$ values.

For $\sigma >1$ and $\a>1$ the function $|f(r)|$ shows a  minimum at $r_{min}=[(\a-1)/(\sigma -1)]^{1/\a}$; the minimum is $y_c=|f(r_{min})|=\a [(\sigma -1)(\a-1)]^{1-1/\a}$
(the radius of a circular caustic). There are two solutions for Eq. (\ref{eq1mod}) if $y>y_c$ and no solutions if $y<y_c$. This case is represented by the dashed
curve (f) in Fig. \ref{sideb}.

\subsection{Analytical solutions}
In this subsection, we list several situations whenever a simple analytical treatment is possible.

The point mass lens model (Schwarzschild lens) is well known, it corresponds to $\a=2$, $r_s=0$, $\sigma=0$. The solutions for the image positions are
$\mathbf{r} = \mathbf{y}(y \pm \sqrt{y ^2 + 4})/2y$ and the total amplification of the two lensed images is  $\mu = (y^2+2)/(y\sqrt{y^2+4})$.

For a single SIDM cored clump, let us assume $\a=2$, $\sigma=0$: critical curves for such a system  exist only when $r_{s}<1$. Under this condition,  two circular critical curves
exist. The first one has a radius $r_z = \sqrt {1 - r_s^2 }$ corresponding to the caustic point $y =0$ (the same as for the point-mass lens), and the second one has a radius $r_c$ such that
$ r_c ^2 = \frac 1 2 \left( {\sqrt {1 + 8r_s^2 } - 1 } \right) - r_s^2 < r_z^2 $ (this expression is positive for $r_{s}<1$),
corresponding to  a caustic with a radius  $y _c  =  r_c \left({1-r_c ^2 - r_s^2  } \right)/\left( {r_c ^2 + r_s^2 } \right)$ ).
In case of a cored clump microlens, the lens equation yields: $r^3 \mp y r^2 + r ( r_s^2 - 1 ) \mp y r_s^2 = 0.$ The roots can be
found using the Cardano method.  For the lower plus, two roots must be taken in the interval $0<r<r_z$ for $y<y_c$, and for the upper minus, one must take the root for $r>r_z$. If $r_{s}>1$
(low-density clump), no caustic exists and only one lensed image can be seen.

For the UCMH model, it is interesting to consider an analytic solution for the overcritical case $\sigma>1$ ($r_s=0,
\a=2$). In this case there exists a critical curve $r_c= 1/\sqrt{\sigma-1}$ and the radius of the corresponding caustic is
described by $y_c= 2 {\sqrt{\sigma-1}}$. Two  images for $y>y_c$ are produced at the positions: $\mathbf{r} = - \mathbf{y}[{y\pm\sqrt{y ^2-4(\sigma-1)}}]/[{2y(\sigma-1)}]. $ The
total amplification of these two lensed images is  $\mu = [y^2-2(\sigma-1)]/[y\sqrt{y^2-4( \sigma-1)}] .$ For $y<y_c$ there are no images and we have an "occultation". The lightcurves
corresponding to this effect, both for point-like and continuous sources, were shown in our previous work \citep{FdPZAS}.

For the case of the cusped lens model, the solutions can be written analytically for $\a=3/2$. The lens equation $ \mathbf {y } = \mathbf {r}\left(1 -  r^{-3/2}\right) $
has two solutions. The first solution must be written separately for $\xi={3 \sqrt{3}}/ {(2 y^{3/2})}<1$ and $\xi>1$:
\[ \mathbf{r} = \frac {\mathbf{y}} 3 \frac {\left[\left(\xi+\sqrt{\xi^2-1}\right)^{2/3} + 1\right]^2} {\left(\xi+\sqrt{\xi^2-1}\right)^{2/3}},\,\, \xi>1;\] and \[
\quad \mathbf{r} = \frac 4 3\, \mathbf{y}\, \cos^2 \left[ \frac 1 3 \arccos \xi\right], \,\, \xi<1.\]

The second solution is \[\mathbf{r} = - \frac {\mathbf{y}} 3 \frac {\left[\left(\xi+\sqrt{\xi^2+1}\right)^{ 2/ 3} - 1\right]^2} {\left(\xi+\sqrt{\xi^2+1}\right)^{2/ 3}}\] for any value of $y$. The magnification curves for these images can be found using the formula (\ref{det}):
$\mu = \left| \left(1  - 1/r^{3/2}\right)\left(1+ 1/(2r^{3/2})\right) \right|^{-1}$.

\section{Comparison between the clump models and the point mass lens}
\label{Compar_models}
\subsection{Observational mimicry of the extended mass microlenses and the source image tracks}
How can one distinguish between the different lens models discussed above on the basis of observations? Of course, the most correct comparison must
come after the fitting of concrete observational data. However, let us first note the qualitative differences  mainly arising from the topological properties of the
corresponding lens mappings, i.e. the  existence of the critical curves, caustics and number of images of a point source. From the above discussion it is easy to see that, for different parameters of the source/lens motion, the caustic intersections can occur leading to appearances/disappearances of the lensed  images.  These events could lead to such observable effects as high  amplifications of the lensed image flux, "occultations"  and sudden jumps of the average image positions (image centroids).

There exist many observational data compiled by EROS, OGLE and other groups \citep{EROS,OGLE} hunting for microlensing events caused by Galactic objects.

Some of these events show a complicated behavior like those characteristic of caustic crossings. This is typically related to the existence of double star systems or planetary systems  \citep{Kains, Shin}. However, most typical HAEs \citep{EROS,OGLE} can be interpreted using the isolated  point mass lens model. It is important to note that there exist some
high amplification events with no detector identified.

On the other hand, for most of the events the deflector-star has been identified; however this case does not rule out the situation
(discussed below) when an  extended clump surrounds  the star. Correspondingly, at this stage we concentrate on the most simple
models of extended microlenses; and we choose the parameters of their motion without  caustic crossings. Namely, we consider
the clump models with two lensed images that will be compared with the "fiducial" model of the usual point mass microlens (the
Schwarzschild lens). For Galactic systems we assume the background continuous matter density to be $\sigma=0$. The critical curves of
the two-image clump models appear to be very similar in case of the Schwarzschild microlens and the cusped clump with $0<\sigma<1$
and $\a>1$. In these cases there is one circular critical curve of unit radius; and also, one caustic point at $y=0$ exists. The case of
a cusped clump with $\a<1$ is topologically different, however it will be also difficult to distinguish its light curve from that
due to the Schwarzschild lens.

\begin{figure}
\centerline{\includegraphics[width=3.0in,height=2.0in]{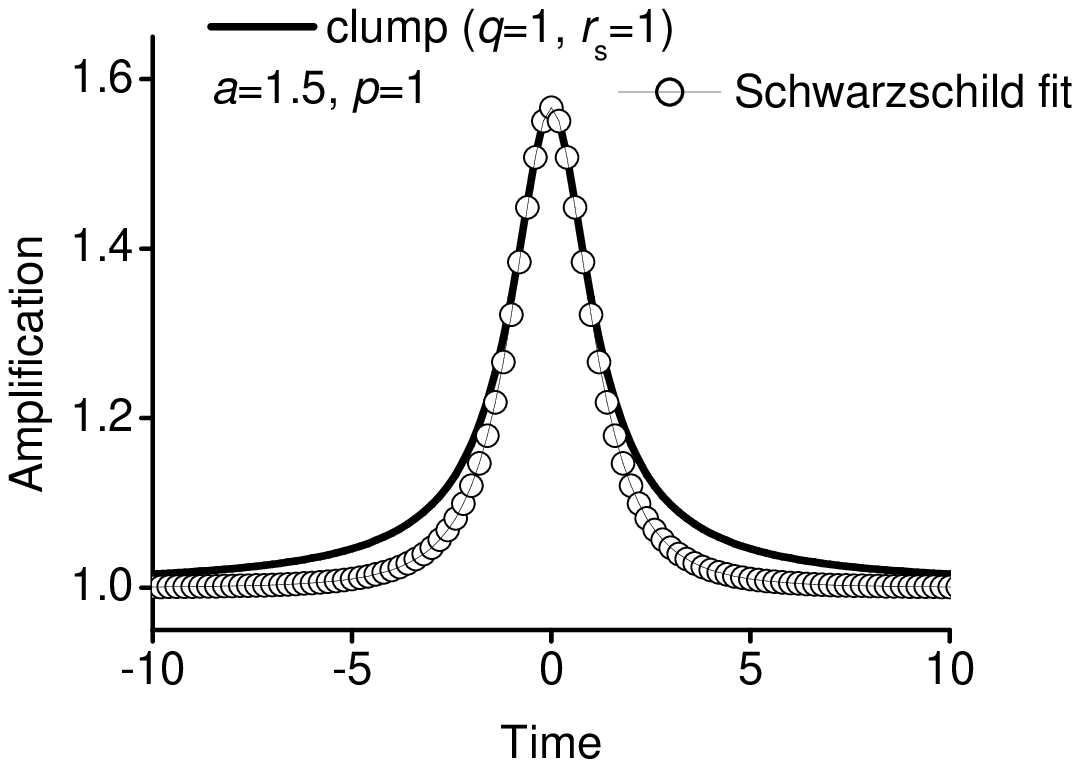}}
\centerline{\includegraphics[width=3.0in,height=2.0in]{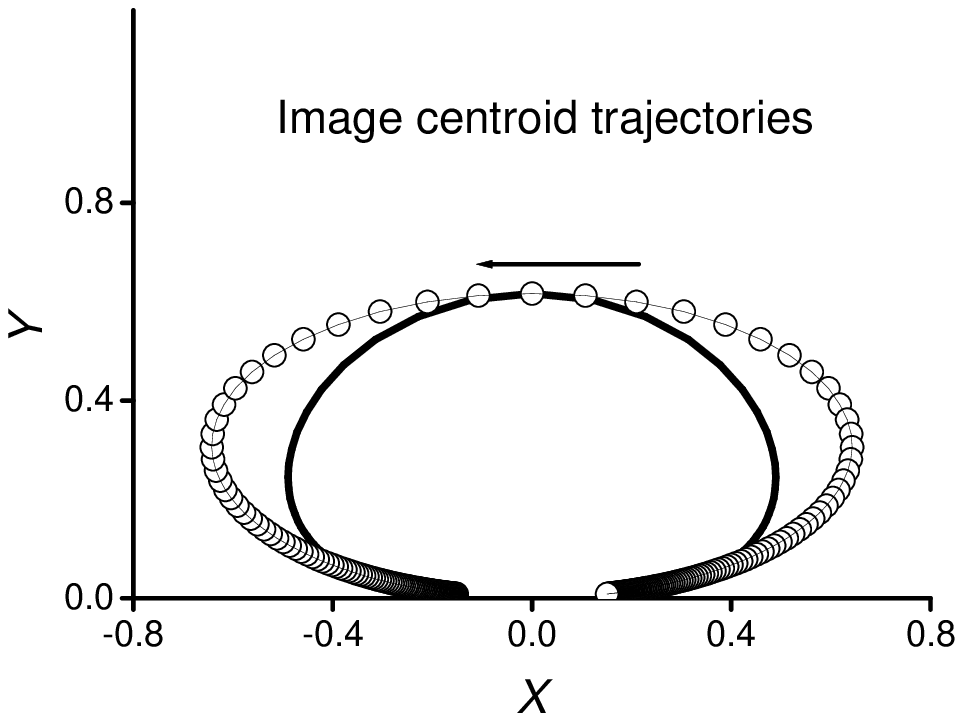}}
\caption{Amplification curves (upper panel) and image centroid
trajectories (lower panel) for the cases of a cored clump (solid) and a point mass
microlens (marked by the small circles). The clump parameters shown in the figure correspond to
Eq. (\ref{p_cl}) ($q=1$ corresponds to a pure clump, $p=1$ is the impact parameter of the clump center with respect to the line-of-sight, its transverse velocity $V=1$). The impact parameter and transverse
velocity of the point mass microlens are derived from fitting
the amplification curve for the clump model. The image centroid trajectory (in the rest frame of the source) in case of the Schwarzshild lens is rescaled in order to clearly indicate the difference between the models.} \label{c-cusp}
\end{figure}

Below we show amplification curves generated with different models corresponding to a straight line motion of the microlensing system with
respect to the line-of-sight towards the remote point source. It turns out that the shape of the amplification curves induced by
two-image clumps are qualitatively similar to the ones induced by a Schwarzschild microlens, and thus they can hardly be
distinguished from one another. Indeed, by an appropriate choice of the impact parameter and the velocity with respect to the line-of-sight, the Schwarzschild lens
light curve can be well fitted by microlensing due to a cusped clump (see Fig. \ref{c-cusp}). For a wide range of  parameters
that may be considered as typical for this problem, the difference between the amplification curves on a plot is  sometimes not
visible to the eye.  One can extract an additional information from the image centroid (IC)\footnote{The image centroid here is a weighted average of positions of all the
images (with the exception of the lenses that we assume to be not visible); the weights are proportional to the amplifications of the
corresponding lensed images.} tracks in the reference frame of the source (see Fig. \ref{c-cusp}, lower panel).  This, however, requires
an accuracy for the positioning measurements at the  microarcsecond level; this is typical for microlensing by Galactic objects and is not accessible at present. E.g., the modern accuracy achieved with HST (WFPC3 camera) is around 20-40 microarcseconds \citep{RCA}.

\begin{figure}
\centerline{\includegraphics[width=3.in,height=2.0in]{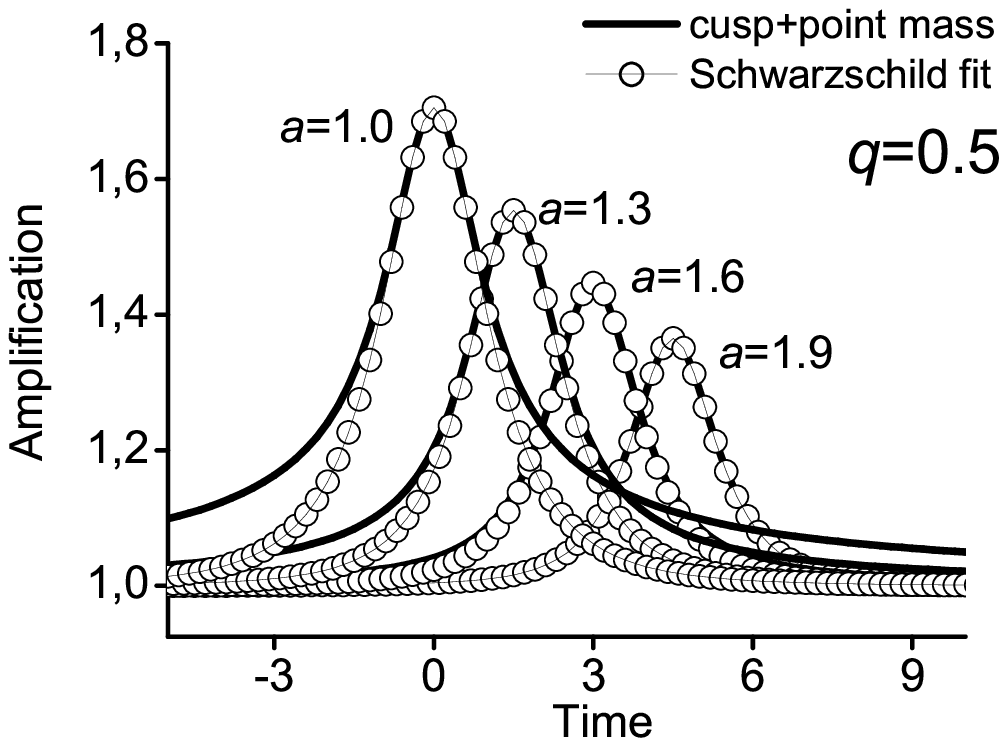}}
\centerline{\includegraphics[width=3.0in,height=2.0in]{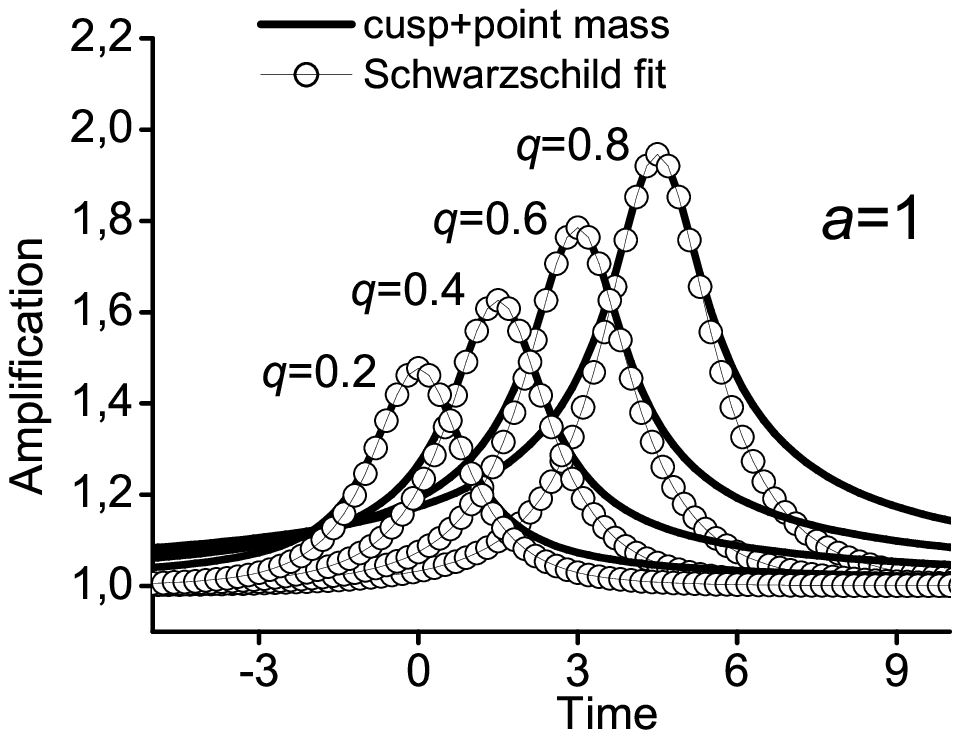}}
\caption{Examples of amplification curves (solid) for microlensing systems that consist of point masses surrounded by cusped clumps with   $q=0.5$, $a=1.0,\,1.3,\,1.6,\,1.9$ (upper panel) $a=1$, and $q=0.2,\,0.4,\,0.6,\,0.8$ (lower panel). The impact parameter of the microlens with respect to the line-of-sight is $p=1$. The curves are fitted by means of point mass microlens models (small circles).  } \label{comp2}
\end{figure}

\begin{figure}
\centerline{\includegraphics[width=3.in,height=2.0in]{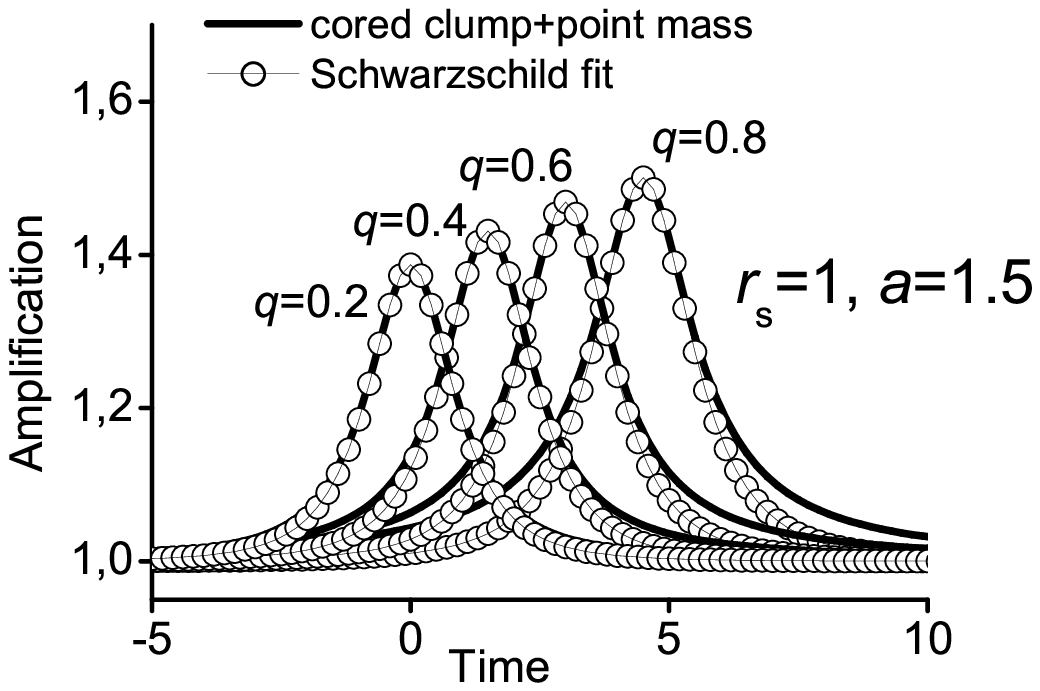}}
\centerline{\includegraphics[width=3.0in,height=2.0in]{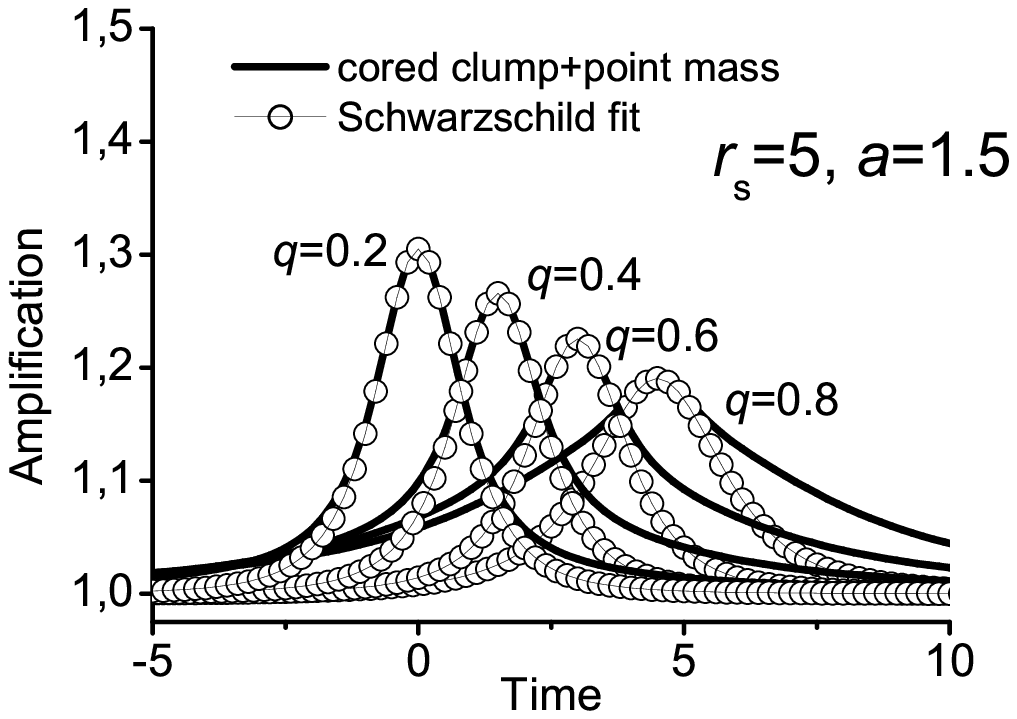}}
\caption{The same as in the lower panel of Fig. \ref{comp2} for cored clump models with $a=1.5$, $r_s=1$ and $r_s=5$. The impact parameter of the microlens with respect to the line-of-sight is $p=1$.} \label{comp3}
\end{figure}

\subsection{Numerical estimates}\label{estimations}

Thus, we proceed in more detail with the differences between the amplification curves corresponding to different microlens models.
Note that we performed a fitting of the amplification curves in Figs. \ref{c-cusp},\,\ref{comp2},\,\ref{comp3}  so as to provide the best
overlapping near the maximum of the curves (where they can be well approximated by a parabola, and where one can expect a good
measurement accuracy). Careful inspection of these figures shows that there is a slight deviation between the curves  in the wings
of the separate HAEs. In this connection, to compare the different microlens models, we proceed as follows.

For a moving extended microlens characterized by an impact parameter $p$  with respect to the line-of-sight, and a transverse
velocity (assumed to be $V=1$), we generate its  amplification curve $\mu(t)$. Then we look for the best fit near the maximum of
this curve using the Schwarzschild lens model light curve ($\mu_{Schw}(t)$); see Appendix \ref{A} for details. Note that as distinct from dealing with the
real observational data we have a simpler problem. In the real case, the fitting parameters, besides the transverse velocity and
the impact parameter of the Schwarzschild lens (which are different from the assumed parameters $p,\,V=1$ of the clump+point
mass complex),  are: the time of maximum amplification and the maximum intensity of the image (or the flux when the lens is
far from the line-of-sight). In our case of the artificial amplification curve, the latter two parameters are fixed.
Thus the difference between the two amplification curves is:
\begin{equation}\label{delta}
\delta=\mathop {\max }\limits_t
\left\{[\mu(t)]^{-1}|\mu(t)-\mu_{Schw}(t)| \right\}.
\end{equation}
This is used to compare the amplification curves due to a somewhat
more complicated model which corresponds to a combination of a point
mass microlens and an extended clump.

\subsection{Microlensing by a point mass and a clump}\label{combination}

It is natural to consider a situation where the cusped or cored DM clump is formed in the same region as where the point mass is situated (and vise versa).  The case of a pure clump microlens  and the Schwarzschild lens represent two extreme cases of this situation. In this connection we consider a microlens model that is represented by the following lens equation:
\begin{equation}
\label{p_cl} \mathbf {y } = \mathbf {r}\left(1- \frac{q}{R^{\a}}-\frac{1-q}{r^2}\right), \quad R=\sqrt{r^2+r_s^2}\, , \end{equation}
where the coefficient $q\,$  describes the relative contribution of the clump, $1-q$ represents that of the point mass lens, and we omitted the contribution due to the background optical depth $\sigma$; $0\le q<1$, $r_s>0$.
The  amplification is $\mu(\mathbf r)=1/D(r)$, where the determinant $D(r)$ is also given by the product $D(r)=\varphi_1(r)\varphi_2(r)$  with
\begin{equation}
\label{factor2a}
\varphi_1(r)=1- \frac{q}{R^{\a}}-\frac{1-q}{r^2},
\end{equation}
\begin{equation}
\label{factor2b}
\varphi_2(r)=1 +
\frac{q(\a-1)}{R^{\a}}-\frac{\a q r_s^2}{R^{\a+2}}+\frac{1-q}{r^2}.
\end{equation}

\begin{figure}
\centerline{\includegraphics[width=2.79in,height=2.38in]{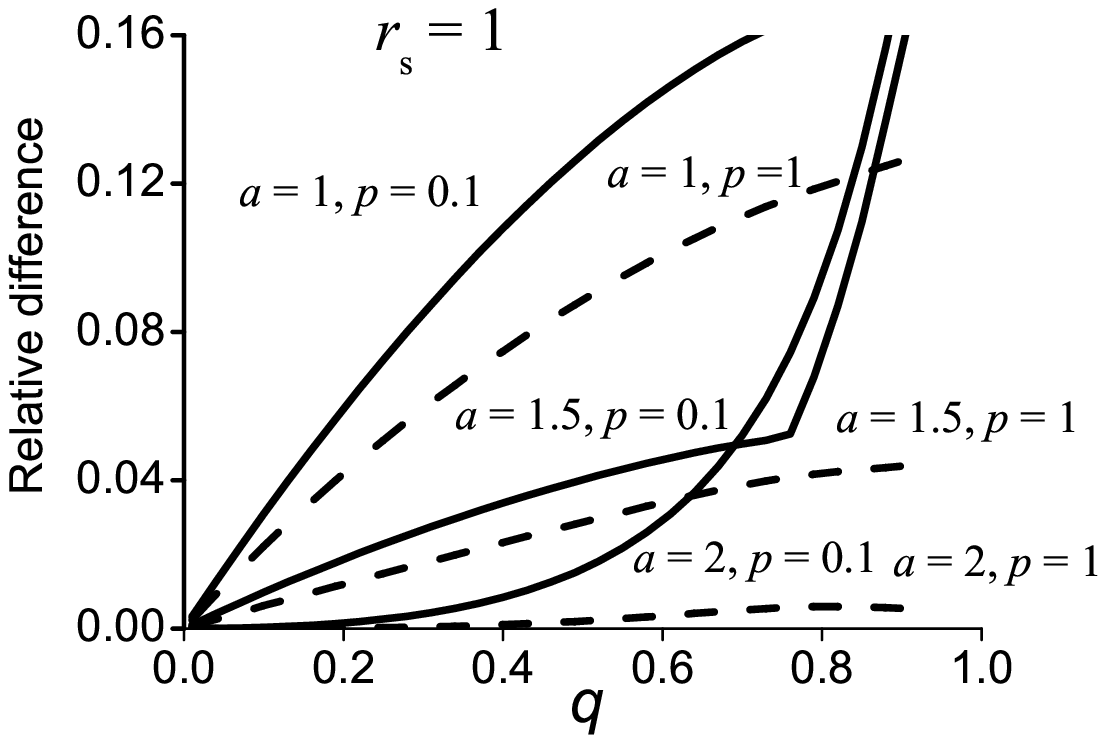}}
\centerline{\includegraphics[width=2.79in,height=2.38in]{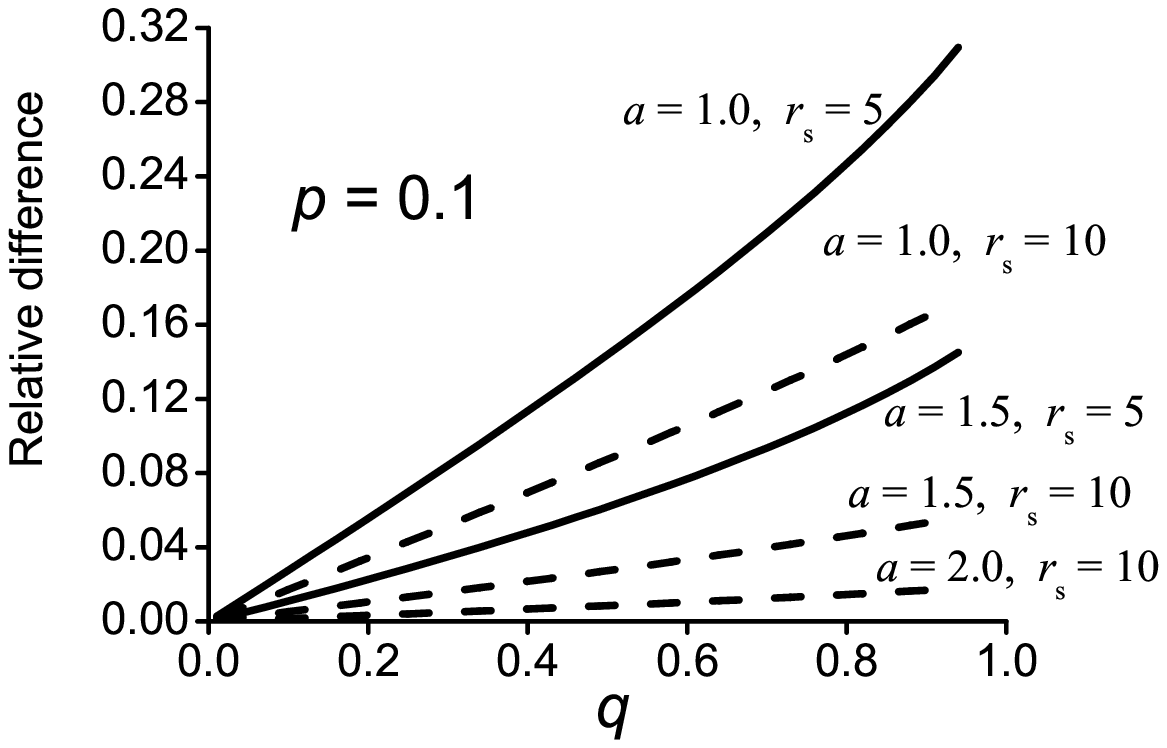}}
\caption{Relative differences $\delta$ of the amplification curves for a different cored clump contributions $q$ in the microlensing
complex "point mass + clump". Upper panel: $a=1.0,\, 1.5,\, 2.0$, impact parameter $p=0.1$ (solid curves), $p=1.0$ (dashed curves),
the clump size $r_s=1$; the amplification is within the limits $\sim 10\div 16$. For $p=0.03$ we have almost the same curves as
for $p=0.1$, but the amplification in this case varies from $\sim 30$ to $50$.  Lower panel: the same values of $a$, $p=0.1$,
$r_s=5$ (solid), $r_s=10$ (dashed).  The curve with $a=2$, $r_s=5$ is superimposed on the curve with $a=1.5$, $r_s=10$ and it is
omitted. Amplification is within $3\div 10$.}\label{Diff_1}
\end{figure}

Furthermore, we assume such numerical values for the parameters of the microlens as to provide the qualitative behavior of the
lens mapping (\ref{p_cl}) to be like that of the Schwarzschild lens with two lensed  images, and so as to provide a considerable
amplification (up to 10 and higher), which can be the best to see the signals of the extended microlens structure. We shall then
estimate the accuracy of measurements needed to differentiate between the lightcurves of these models. The dependence of this
difference upon the contribution $q$ of the clump in the lensing complex "cored clump + point mass" is shown in Fig.
\ref{Diff_1} for some values of $\a$ and $r_s$.

\begin{figure}
\centerline{\includegraphics[width=2.79in,height=2.38in]{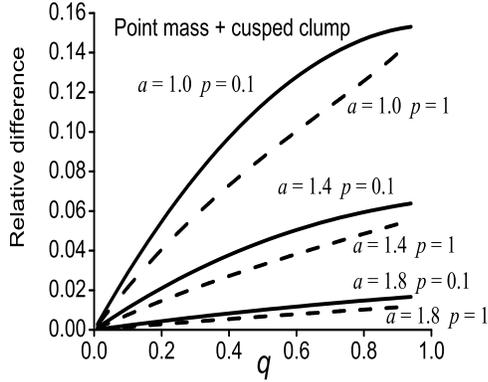}}
\caption{Relative differences $\delta$ of the amplification curves for different cusped clump contributions $q$ in the microlensing complex "point mass + clump". Here $a=1.0,\, 1.5,\, 1.8$. For the impact parameter $p=0.1$ (solid) the amplification is within $\sim 10\div 20$; for $p=1$ (dashed) the amplification is $\sim 1.3\div 2$. For $p=0.03$ we have practically the same curves as for $p=0.1$, but the amplification varies from $\sim 30$ to $60$.}\label{Diff_2}
\end{figure}

In case of the configuration of a point mass and a cusped clump we assume $q\in (0,1)$, $r_s=0$. The relative differences between the
amplification curve of this complex and that of the fitted Schwarzschild lens are shown in Fig. \ref{Diff_2}.

\section{Statistical effects of DM clumps microlensing in extragalactic GLS}\label{statistical}

In a typical extragalactic GLS we have several macro-images of one quasar. Comparison of the amplification curves of these images
makes it possible to separate the proper brightness variations of the quasar and to derive variations due to the microlensing
processes. Contrary to the case of the Galaxy, because of  the considerable microlensing optical depth in extragalactic systems,
instead of an isolated point mass or putative DM clump, we must deal with an aggregate of masses in the lensing galaxy. In this
case we have a complicated caustic network generated by unknown masses with unknown positions. Therefore, the problem takes a
statistical shade \citep{PaczWambs,WambPaKa,ScW}. Observationally, the main source of information in a concrete GLS is at present an
effective microlensed light curve of every image (i.e., separated from proper brightness variations that are intrinsic to the
source), which arises because of the source motion\footnote{which is again assumed to be a uniform straight line motion.}. As
before, for a theoretical treatment we deal with the amplification curves, that is the dependence of the amplification coefficients
upon the time.

Our aim is to estimate the statistical effect of the extended masses (clumps) on the autocorrelation functions of the amplification
curves. To compare the amplification curves in microlensing systems with a different content of these clumps, we consider a simple model of
stochastic point and extended masses. We confine ourselves to a special case of the cored clumps according to Eq. (\ref{eq1z})
with $a=2$ (cf. \cite{ISRN_Zh}). An external shear owing to the average field of the lensing galaxy will also be taken into account. Thus, we use the
lens equation for N masses in one lens plane
\begin{equation}
\label{lens_eq_clumps} {\mathbf y}= \hat{A} {\mathbf r}- \sum_{i = 1}^N
\frac{R_{E,i}^2({\mathbf r}-{\mathbf r_i})}{|{\mathbf r}-{\mathbf r_i}|^2 +r_{s,i}^2}
\end{equation}
\noindent where $\hat{A}=diag\{1-\gamma\,,\,\, 1+\gamma\}$ is the two-dimensional external shear matrix (the optical depth of the background continuous matter is taken to be zero), $r_{s,i}$ is a characteristic  size of the $i$-th cored clump with mass $M_i$, ${\mathbf r_i}$ is the position of its center on the lens plane, and $R_{E,i}$ its Einstein ring radius:
$R_{E,i}^2= 4G M_i D_{ds}/(c^{\,2} D_d D_s)$. Obviously, for $r_{s,i}=0$ we have ordinary point  microlenses with mass $M_i$.

In  our simulations, the microlens masses and positions were chosen in a random way. The positions ${\mathbf r_i}$  were
distributed homogeneously inside a circle, which was big enough to minimize the boundary effects; also, to check the result
convergency we considered different sizes of the circle. The mass distribution followed the Salpeter's law \citep{salpeter} with a
power-law index $-2.35$ within the mass range $M_{i} \in [0.4; 10] M_{\odot}$. In every set of numerical experiments, the input
parameters (that were controlled to be the same in all the realizations of the set) were as follows: the total optical depth
of the microlensing field $\sigma_{tot}$, the optical depths $\sigma_{p}$ ,  $\sigma_{cl}$ of the point masses and the extended
clumps ($\sigma_{cl}+\sigma_{p}=\sigma_{tot}$), and the relative size of the clumps $\kappa=r_{s,i}/R_{E,i}$. To introduce the
clumps into the microlensing field, we replaced  some randomly chosen part of point masses by extended clumps without changing
their positions $r_{i}$, masses $M_i$; the size parameter was $r_{s,i}=\kappa R_{E,i}$. Then, for each realization of the
microlensing field we obtain an amplification map by means of the ray shooting method combined with our GPU-enabled microlensing C++ code based on the hierarchical tree algorythm
\citep{SchnEhlFal, Wambs}. Convolution of the amplification map with the normalized brightness distribution over the initial source
$I({\mathbf y})$ yields the amplification coefficient $\mu$. The mathematical representation of this procedure can be written as \citep{Alex,ISRN_Zh}
\[
\mu({\mathbf y})=\int \int I[{\mathbf Y}({\mathbf x})-{\mathbf y}] dx_1 dx_2,
\]
where ${\mathbf y}$ is the position of the source center in the source plane projected in the lens plane and ${\mathbf y}=
{\mathbf Y}({\mathbf r})$ represents the lens mapping ${\mathbf r}\to {\mathbf y}$. We used a Gaussian model for $I({\mathbf
y})$; the source half-brightness radius was $R_{1/2}=0.2$, its motion has been assumed to be uniform. Due to the motion of the
source we have an amplification curve $\mu(t)$.

Having a large number of  realizations of the microlensing  field and the corresponding amplification curves $\mu(t)$, we have calculated the ACFs
\begin{equation}\label{autocorr}
A(\tau)=(\Delta \mu)^{-2} <[\mu(t)-\mu_0][\mu(t+\tau)-\mu_0]>,
\end{equation}
where
\begin{equation}\label{average}
\mu_0=<\mu(t)>,\quad \Delta \mu=\sqrt{<[\mu(t)-\mu_0]^2>},
\end{equation}
the brackets $<...>$ denote the averaging over all the realizations of the amplification curves for the fixed  value of the optical depth of continuous masses $\sigma_{cl}$ and point masses $\sigma_{p}$.

To consider the possible observational manifestations of the extended microlenses, their size  $R_c$ has been chosen to be
comparable with a typical Einstein ring radius $R_E$ of the microlenses; otherwise in the limiting situations we approach to
either the well-known case of a continuous background matter for $R_c>>R_E$, or we have a standard microlensing of point masses.
Here we present the results for  $\sigma_{tot} = 0.3$, and $\kappa=5$. The parameter $\sigma_{cl}$ has been varied from zero
to $0.3$. The length unit corresponds to $R_{E,\odot}= [4G M_{\odot} D_{ds}/(c^{\,2} D_d D_s)]^{1/2}=1$. Furthermore, $t$ and
$\tau$ are in units of $R_{E,\odot}$ (i.e. assuming a velocity $V=1$);  the normalization to $M_{\odot}$ is just
conventional.

\begin{figure}
\vskip-3mm\includegraphics[width=80mm,height=70mm]{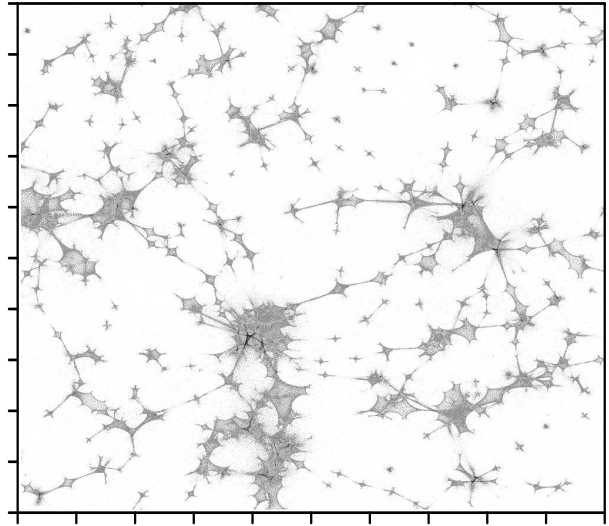}
\vskip-3mm\caption{Amplification map for $\sigma_{cl} = 0, \gamma=0, \, \sigma_{tot}= 0.3$. } \label{map_1}
\end{figure}

\begin{figure}
\vskip-3mm\includegraphics[width=80mm,height=70mm]{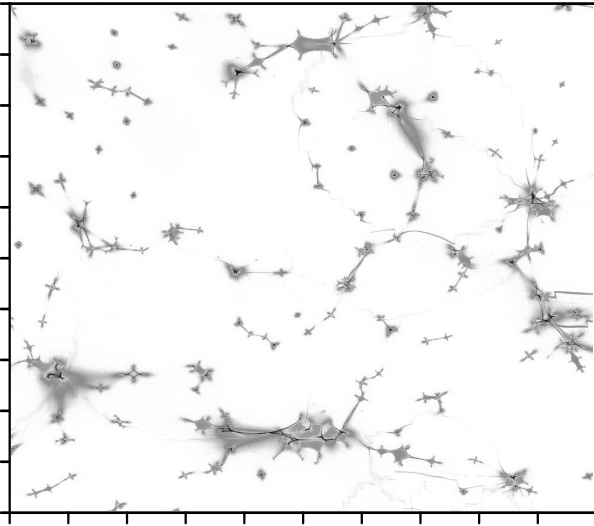}
\vskip-3mm\caption{The amplification map for $\sigma_{cl} = 0.2, \gamma=0$, here and below $\sigma_{tot}= 0.3$, and the size parameter is $\kappa = 5$.}\label{map_2}
\end{figure}

\begin{figure}
\vskip-3mm\includegraphics[width=80mm,height=70mm]{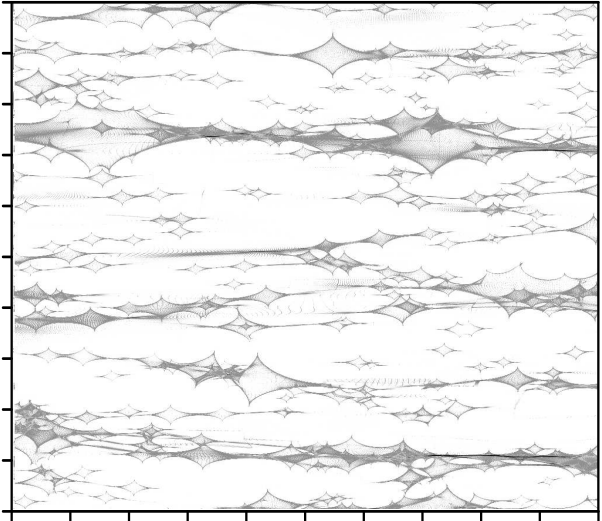}
\vskip-3mm\caption{The amplification map for $\sigma_{cl} = 0.2, \gamma=0.5$.}\label{map_3}
\end{figure}

\begin{figure}
\vskip-3mm\includegraphics[width=80mm]{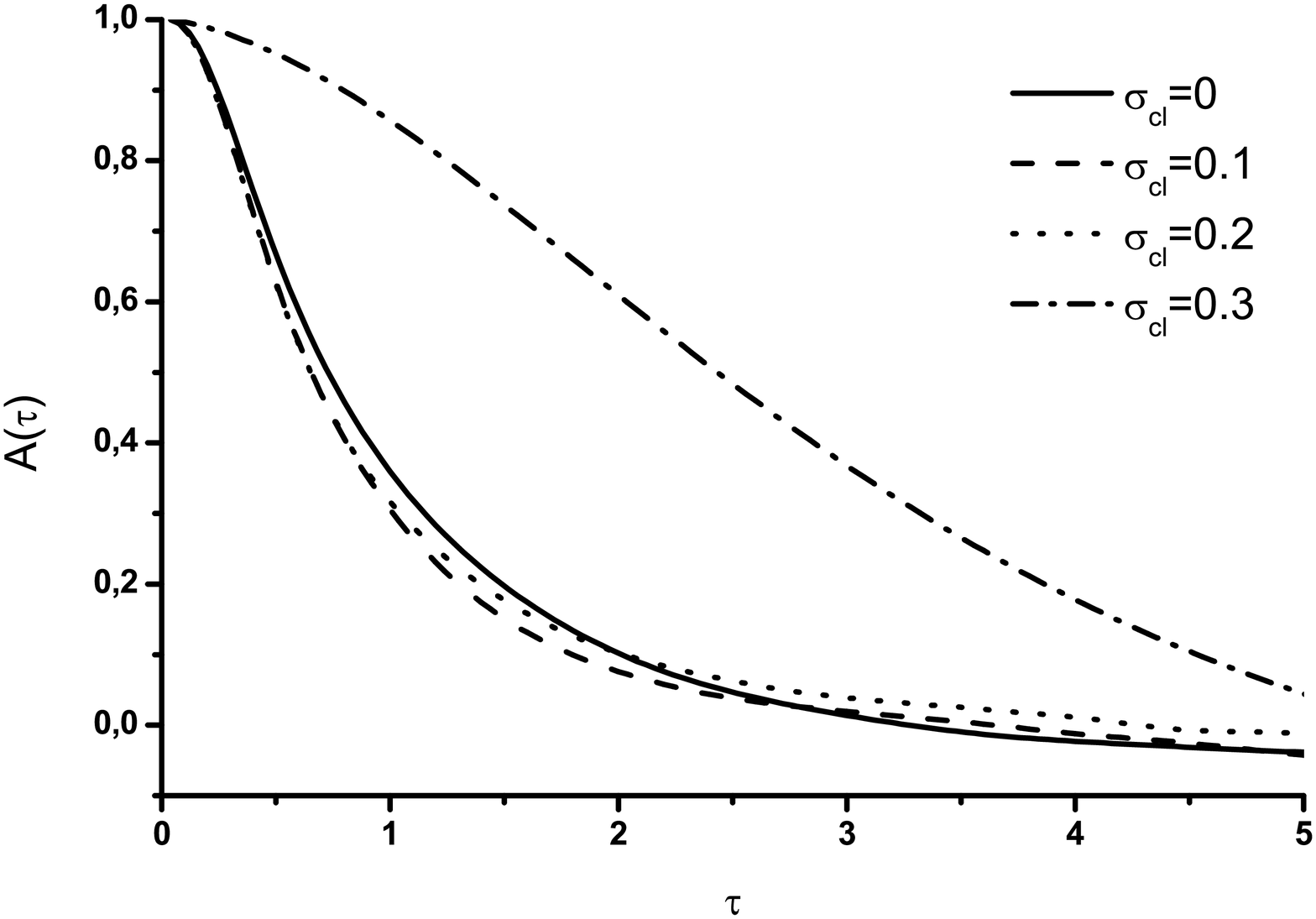}
\vskip-3mm\caption{ACFs of the amplification curves for the case  $\sigma_{tot}= 0.3$, $\kappa=5$, $\gamma=0$ and for $\sigma_{cl}=0,0.1,0.2,0.3$.}\label{autocor}
\end{figure}

\begin{figure}
\vskip-3mm\includegraphics[width=80mm]{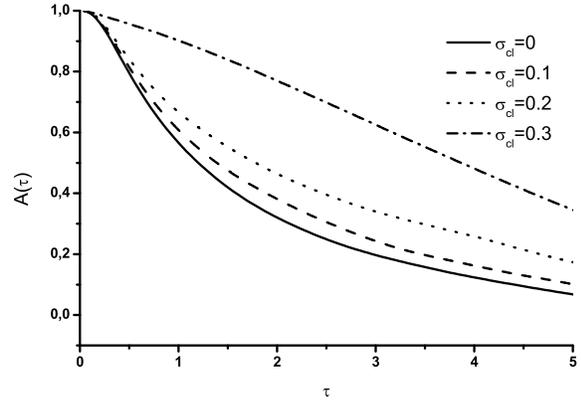}
\vskip-3mm\caption{ACFs of the amplification curves for the case   $\sigma_{tot}= 0.3$, $\,\kappa=5$, $\gamma=0.5$ and for $\sigma_{cl}=0,0.1,0.2,0.3$. }\label{auto-gamma}
\end{figure}

The examples of the amplification maps in the source plane are shown in  Figs. \ref{map_1} - \ref{map_3}. The amplification distributions for various fractions of clumps (and the same $\sigma_{tot}=0.3$, $\gamma=0$) noticeably differ one from another. As can be expected, the presence of the clumps makes this distribution  smoother. The presence of non-zero $\gamma$ (Fig. \ref{map_3}) stretches the maps in the direction of the shear. For every set of parameters $\sigma_{tot},\sigma_{cl},\gamma $ we generated typically several hundred maps and the corresponding  amplification curves.

Using the ensemble of these curves, the averaging procedure yields ACF  $A(\tau)$. These functions are shown in Figs. \ref{autocor}, \ref{auto-gamma}
for different fractions of point masses and clumps. We note that the behavior  of $A(\tau)$ and $\Delta\mu$ as functions of $\sigma_{cl}$ appears to be rather  complicated.
E.g., for $\gamma=0$ we see a non-monotonic variation of the curves as $\sigma_{cl}$ changes (Fig. \ref{autocor}).
This effect is not observed for non-zero values of the shear (see the case $\gamma=0.5$   in Fig. \ref{auto-gamma}).

\section{Discussion}

We have analyzed the observational signatures of gravitational microlensing in presence of extended  masses (clumps), which can represent, for example,
hypothetical structures arising in cold DM models as a result of clustering. To produce  significant effects due to the presence of the clumps, the masses
of the clumps were chosen of the same order as those of the point masses, and the sizes of the clumps were of the  order of several typical Einstein radii.
We compared the artificial amplification curves (essentially, the light curves) arising due to a relative motion of the microlens with respect to a
remote source.

We have proposed simple models that describe  circular symmetric cusped or cored clumps and their combinations with point masses.
We do not consider these idealized models as the only possible ones, rather, our aim was to look at some "toy problems" in order
to see the main qualitative features accompanying the microlensing process in our Galaxy and beyond in presence of  small-scale DM
clustering, and to estimate the order of the effects involved. We believe that our main findings on this matter will be preserved at
a qualitative level for more complicated models of the DM clumps, though an exact final answer will be the subject of a separate
investigation.

In case of Galactic microlensing, the case of isolated circular symmetric microlenses of Section \ref{extended_clump_models}
has led to a detailed analytical treatment; here the regions of parameters of the lens mapping yielding the occurrence of one, two
or three images were  determined. We note that any  structures yielding  changes in the number of images (cf. caustic crossings,
occultations) are possible  only in case of a considerable local dark matter density, either near the clump centers or around
the point masses. The most vivid example is an eclipse-like effect with the total disappearance of the lensed images. Though, we do
not believe this case to be  realistic for the Galaxy conditions, we cannot exclude it at all. It may be difficult to distinguish
such complicated events from those related to double stars or planetary systems; here a detailed investigation of the light
curves of concrete high amplification events (HAEs) is required.

We considered relatively simple models and events. We found that the light curve of a HAE in presence of the extended microlens (the
isolated clump or its superposition with the point mass)  can be well mimicked by a light curve of the Schwarzschild  lens with
appropriately chosen mass and parameters associated with its motion. Nevertheless, in case of such a fitting, there remain some
residuals in the wings of the HAE light curve that can be used to distinguish between the microlens models. These residuals have
been estimated through the fitting of the light curves of our clump model by those of the point mass lens (Sections
\ref{estimations}, \ref{combination}); these results determine the required level of photometric accuracy needed to detect the
signals from the extended clumps. This level can be estimated from Figs. \ref{Diff_1}, \ref{Diff_2}. In particular, for cored clump +
point mass system with $r_s=p=1$ (that is, the core size and impact parameter are equal to the Einstein ring radius) and average
values $q=0.5, a=1.5$  the light curve deviation from that of the ordinary point mass lens can be detected if the photometric accuracy is at the $\sim
0.02\, mag$  level. So is also the case of the cusped clump, and the requirements to the accuracy in case of a pure cusped or cored
clump ($q=1$) are even lower.

We also note that the degeneracy in the choice of the model can be reduced  by simultaneous astrometric observations of the source
image centroid motion (see Fig. \ref{c-cusp}). If this information is available, this may help to restrict the parameters of the
microlensing system. Though this does not mean that all of them can be determined uniquely, not to mention the question of
uniqueness of the microlensing model itself.

In any case, there is a region of the clump parameters that induces a considerable difference  between the light curves, which
is within the reach of modern photometric measurements. Therefore, we see that at least some extended microlens models can be, in
principle, either confirmed or rejected from the use of high-quality observations. However, in real observations of microlensing by Galactic objects, the situation will be aggravated by the possible presence of unobservable planets and star companions. Then in this case a more detailed treatment of concrete high amplification events is also necessary to compare the models that involve
binaries and planets with those dealing with extended clumps.

 In extragalactic GLSs the situation is much more complicated. The view of amplification maps and the caustic network generated by simple point mass microlenses
 (i1) seems to be rather different from those in presence of a considerable contribution of extended clumps (i2).
 Nevertheless, individual light curves in case (i1) can also include high amplification and de-amplification effects which have an analogous  look as in case (i2).
 The other complication in  extragalactic GLSs is that owing to the non-negligible source size since both the light curves and the centroid brightness trajectory
 will be smoothed out significantly.  To have an idea about microlensing in cases (i1) and (i2),
 we considered statistical models of many-body microlensing systems that involve both point  and extended microlenses.  The latter have been represented by a special case of the cored clumps.

Considerations of Section \ref{statistical} have much in common with considerations by \cite{Metcalf,Chiba,Dal-Koch,ScW} that, among
other questions, deal with the issue of microlensing by extended structures. In particular, \cite{ScW} pay attention to microlensing brightness
variations in presence of a variable smooth matter depending on the parity of the close macrolens images. These papers, however, are mainly aimed at investigating the problem of anomalous
flux ratios and therefore they deal with  other mass and spatial scales.  Our simulations deal with
extended microlenses (e.g., DM clumps) of  stellar mass, with sizes comparable to that of corresponding Einstein rings -- in order to have a noticeable effect in the light curves.
A technical difference of our simulations owes to the simplified model of the lens mapping in presence of the clumps, which enables us to speed up
the numerical calculations of ACFs with a rather  modest computer hardware.  For every choice of parameters that characterize the content of the clumps, their sizes
 and masses, we generated an ensemble of  amplification curves assuming a uniform spatial distribution of microlenses in the lens plane.
 Using these curves we built ACFs of the amplification curves for a set of optical depths of  extended clumps and  point masses. Joint use  of microlensed flux variations and ACFs may provide a means for determining the clump fraction and the other parameters of the lensing system.

Indeed, there is a notable difference between the ACFs for  different values of the clump contribution. In this view, we note the remark by  \cite{ScW} that the microlensing variations
appear to be "less dramatic for extended subhalos than for point microlenses". This also has been mostly observed in our simulations, though with some exceptions.
One could expect that by and large the extended structures  make ACFs less steep and therefore, this effect shall increase the corresponding correlation length, and this typically is the case for  $\sigma_{cl}\le 0.3$ and $\gamma\in[0,0.5]$  ($\sigma_{tot}=0.3$) ), but also with some exceptions showing a complicated dependence of ACFs upon $\sigma_{cl}$ (Figs. \ref{autocor}, \ref{auto-gamma}).

However, detection of these effects requires sufficiently long observations of the light curves in real GLSs. The relevance of
our results to   extragalactic systems   is connected also with the occurrence of a significant optical depth $\sigma_{cl}$. This
depends on the surface mass density along the line-of-sight which is not well known. For very rough estimations one can use the
masses and sizes of the galactic haloes as compared to those of the stellar population \footnote{see, e.g., the examples from
\cite{Shull}}. The density of DM in the Universe is five times higher than the baryonic one and the percentage of the luminous
matter in galaxies is even lower  by one order. Then, it may be quite possible to have the surface mass density of the extended
clumps of some GLS of the same order as that of the  point masses (stars). Moreover, despite the mass-luminosity ratio for typical
galaxies is known to be at the level of $M/L \sim 2 - 4$, for some spherical dwarf galaxies it is $M/L \sim 100$ \citep{Swaters}.

\section{Conclusions}\label{conclusions}
To sum up, we studied  microlensing of a remote source by point and extended masses, the latter representing hypothetical results
of cold DM clustering. In case when the masses of the extended clumps are comparable to those of the point objects (stars) and
their characteristic sizes are of the same order as the Einstein radius of the stars, our findings are as follows:
\begin{itemize}
\item We proposed a simple circular symmetric model of microlensing by isolated dark matter clumps that allowed us a detailed investigation of the corresponding lens mapping.
\item In case of a single point source motion, we built amplification curves of the lensed images; this is done in case of microlensing by the isolated clumps
and/or concentric clumps and point masses. For a wide range of model parameters, we estimated the residuals after fitting these amplification curves with those
due to the single point mass microlensing.
\item  The results allow one to estimate the photometric accuracy needed to differentiate the model with the extended clumps and without them. We note that a number of model parameters can be ruled out under the level of photometric accuracy $\Delta m=0.01--0.02$.
\item In case of extragalactic gravitational lens systems, we built autocorrelation functions of the amplification curves for different clump contributions.
We found a noticeable difference between ACFs in case of a considerable value for the clump contribution and without the clumps.
\end{itemize}

\section{Acknowledgements}

We thank the anonymous referee for her/his helpful comments. EF acknowledges the AGO/AEOS Institute of Astrophysics and Geophysics of Li\`ege University for their warm hospitality and
providing the technical facilities for this work. EF and JS acknowledge support from the ESA PRODEX programme ``Gaia-DPAC'', and from the
Belgian Federal Science Policy Office. VIZ acknowledges partial support from the State Fund for Fundamental Research of Ukraine, project $\Phi$64/42-2015. This work has been supported in part by the scientific program ``Astronomy and Space Physics",  Taras Shevchenko National
University of Kyiv.

\label{lastpage}
\appendix
\section[A]{Fitting the amplification curve in case of a Schwarzschild lens}
\label{A}
In case of a point mass (Schwarzschild) lens we have the mapping
\begin{equation} \mathbf {y } = \mathbf {r}\left(1-\frac{R_E^2}{r^2} \right)
\end{equation}
The amplification is well known:
\begin{equation}\label{schwarzschild}
\mu (\mathbf {y })=\frac{y^2+2 R_E^2}{y\sqrt{y^2+4 R_E^2}}\,\,.
\end{equation}
In case of a relative straight-line motion of the lens and the source
we have $y^2/R_E^2=p_S^2+V_S^2t^2$, $V_S$ is the velocity, $p_S$ is the impact parameter. Therefore, the problem of fitting any given dependence $\mu(t)$ (e.g., that has been  obtained for a clumped or cusped lens model) by means of (\ref{schwarzschild}) is reduced to  finding the coefficients $a,b$ of an approximate linear mapping $t^2\to y$:
\begin{equation}
a+bt^2 =2\left(\frac{\mu(t)}{\sqrt{\mu^2(t)-1}}-1\right),
\end{equation}
where $a=p_S^2,\, b=V_S^2$.
Obviously, $p_S$ and $V_S$ will differ from that of the initial clump model.
 In our case, to estimate the difference between models, the approximation is performed near the maximum of the amplification curve, where it has a parabolic form (typically for $|t|< 0.05$. Here residuals of the approximation are of the order of $10^{-5}\div 10^{-8}$). Using the point mass model with the parameters obtained as the result of this approximation, we estimated $\delta$ for a larger interval.

Note that to determine $p_S$ and $V_S$  we deal with the relative values $y/R_E$ irrespective of $R_E$, which is not essential for fitting the amplification. However, this is essential when considering the image centroid. E.g., in case of two images, the image centroid is
\[
\mathbf{r_c}=\frac{\mu_1\mathbf{r_1}+\mu_2\mathbf{r_1}}{\mu_1+\mu_2},
\]
where $\mathbf{r_1}$ is the $i$-th image position and $\mu_i$ is its amplification ($i=1,2$).
Here we have an additional parameter $R_E$ that can be used to rescale the trajectory $\mathbf{r_c}(t)$; this is done in the lower panel of Fig. \ref{c-cusp}.


\begin{thebibliography}{99}

\bibitem[\protect\citeauthoryear{Ade et al.}{2013}]{Ade} Ade P. A. R. et al. 2013  A\&A \textbf{1303}, 5062.

\bibitem[\protect\citeauthoryear{Alcock et al.}{1993}]{Alcock} Alcock C., Akerlof C. W., Allsman R. A., et al., 1993, Nat, \textbf{365}, 621.

\bibitem[\protect\citeauthoryear{Alexandrov \& Zhdanov}{2011}]{Alex} Alexandrov A.N.,  Zhdanov V.I., 2011, MNRAS, \textbf{417}, 541.


\bibitem[\protect\citeauthoryear{Aubourg et al.}{1993}]{Aubourg} Aubourg E.; Bareyre P.; Br\'ehin S.  E. et al., 1993, Nat, \textbf{365}, 623.

\bibitem[\protect\citeauthoryear{Baryshev \& Bukhmastova}{1997}]{BB} Baryshev Yu., Bukhmastova Yu. 1997, Astron.Rep., \textbf{41}, 436.

\bibitem[\protect\citeauthoryear{Berezinsky et al.}{2008}]{BDE} Berezinsky V., Dokuchaev V., Eroshenko Yu. 2008, Phys.Rev.D, \textbf{77}, 083519.

\bibitem[\protect\citeauthoryear{Berezinsky et al.}{2013}]{BDE2} Berezinsky V., Dokuchaev V., Eroshenko Yu. 2013, JCAP, \textbf{11}, 059.


\bibitem[\protect\citeauthoryear{Brada\v c et al.}{2006}]{Bradac_2006} Brada\v c M., Clowe D., Gonzalez A. H. et al. 2006, ApJ, \textbf{652}, 937.

\bibitem[\protect\citeauthoryear{Chen et al.}{2007}]{C} Chen J. et al. 2007, ApJ, \textbf{652}, 52.

\bibitem[\protect\citeauthoryear{Chiba}{2002}]{Chiba} Chiba M. 2002 ApJ, \textbf{565}, 17.

\bibitem[\protect\citeauthoryear{Clowe et al.}{2004}]{Clowe_2004} Clowe D., Gonzalez A., \& Markevitch M. 2004, ApJ, \textbf{604}, 596

\bibitem[\protect\citeauthoryear{Clowe et al.}{2006}]{Clowe_2006} Clowe D., Brada\v c M., Gonzalez A. H. et al. 2006, ApJL, \textbf{648}, L109.

\bibitem[\protect\citeauthoryear{Congdon et al.}{2010}]{CKN} Congdon A., Keeton C., Nordgren C. 2010, Aph.J., \textbf{709}, 552.

\bibitem[\protect\citeauthoryear{Contini et al.}{2011}]{CLB} Contini E., De Lucia G., Borgani S. 2012, MNRAS, \textbf{420}, 2978.


\bibitem[\protect\citeauthoryear{Dalal \&  Kochanek}{2002}]{Dal-Koch}  Dalal N.,  Kochanek C. S. 2002, ApJ, \textbf{572}, 25.

\bibitem[\protect\citeauthoryear{Del Popolo}{2014}]{DP}  Del Popolo A. 2014, Int.Journ. Mod. Phys. D, \textbf{23}, N3, 1430005.

\bibitem[\protect\citeauthoryear{Del Popolo et al.}{2014}]{DLF} Del Popolo A., Lima J. A. S., Fabris J. C., Rodrigues D.C. 2014, JCAP, \textbf{4}, id.21

\bibitem[\protect\citeauthoryear{Diemand et al.}{2004}]{DMS} Diemand J., Moore B., Stadel J. 2004, MNRAS, \textbf{352}, Is. 2, 535.

\bibitem[\protect\citeauthoryear{Diemand et al.}{2005}]{D2} Diemand J., Moore B., Stadel J. 2005, Nature, \textbf{433}, 389-391.

\bibitem[\protect\citeauthoryear{Dobler et al.}{2007}]{Dobler2007} Dobler G., Keeton C.R.,  Wambsganss J. 2007, MNRAS, \textbf{377}, 977-986.


\bibitem[\protect\citeauthoryear{Fedorova et al.}{2014}]{FdPZAS} Fedorova E., Del Popolo A., Zhdanov V.I., Alexandrov A.N., Sliusar V. 2014,
Rencontres de Moriond 49th. Cosmology, Eds: E.Auge, J.Dumarchez
and J. Tran Thanh Van al. La Thuile 2014 Proceedings, 407.

\bibitem[\protect\citeauthoryear{Jeans}{1923}]{Jeans} Jeans J. 1923, MNRAS, \textbf{84}, 60.

\bibitem[\protect\citeauthoryear{Hisano et al.}{2006}]{HIT} Hisano J., Inoue K.T., Takahashi T. 2006, Phys.Lett.B., \textbf{643}, 141.

\bibitem[\protect\citeauthoryear{Kains et al.}{2013}]{Kains} Kains, N.; Street, R. A.; Choi, J.-Y.; Han, C.; Udalski, A. et al. 2013, A\&A, \textbf{552}, id.A70, 10 pp.

\bibitem[\protect\citeauthoryear{Kapteyn}{1922}]{Kap} Kapteyn J.C. 1922, Aph.J. \textbf{55}, 302.

\bibitem[\protect\citeauthoryear{Keeton \& Moustakas}{2009}]{KM} Keeton C.R., Moustakas L.A. 2009, ApJ, \textbf{699}, 1720.

\bibitem[\protect\citeauthoryear{Klypin et al.}{1999}]{Klypin} Klypin A., Kravtsov A., Valenzuela O., Prada F. 1999, Aph.J., \textbf{522}, 82.

\bibitem[\protect\citeauthoryear{Knebe et al.}{2008}]{KAP} Knebe A., Arnold B., Power C., Gibson B.K. 2008, MNRAS, \textbf{386(2)}, 1029.

\bibitem[\protect\citeauthoryear{Knebe et al.}{2003}]{KDGS} Knebe A., Devriendt J., Gibson B., Silk J. 2003, MNRAS, \textbf{345}, 1286.

\bibitem[\protect\citeauthoryear{Kormendy \& Freeman}{2003}]{K} Kormendy J, Freeman K.C. 2003, IAU Symp.220 Eds: S.D.Ryder et al. San Francisco: Astron. Soc. Pacific., 377; Astro-Ph/0407321.

\bibitem[\protect\citeauthoryear{Lee et al.}{2009}]{LAK} Lee S.K. Ando A., Kamionkowski M. 2009, JCAP, \textbf{07}.

\bibitem[\protect\citeauthoryear{Markevitch et al.}{2003}]{Bullet}Markevitch M., Gonzalez A., Clowe D. et al. 2003, Aph.J., \textbf{606}, 819.

\bibitem[\protect\citeauthoryear{MacLeod et al.}{2012}]{ML} MacLeod C. et al. 2013, AphJ., \textbf{773}, 35.

\bibitem[\protect\citeauthoryear{Mao et al.}{2004}]{MJOW} Mao S., Jing Y., Ostriker J. P., Weller J. 2004, ApJ, \textbf{604}, L5.

\bibitem[\protect\citeauthoryear{McKean et al.}{2007}]{MK} McKean J. P. et al. 2007, MNRAS, \textbf{378(1)}, 109.

\bibitem[\protect\citeauthoryear{Metcalf \& Madau}{2001}]{Metcalf} Metcalf R.B., Madau P., 2001, ApJ \textbf{563}, 9.

\bibitem[\protect\citeauthoryear{Metcalf et al.}{2003}]{MMM} Metcalf R.B., Moustakas L. A. Bunker A. J., Parry I. R. 2003, Aph.J. \textbf{607(1)}, 43.

\bibitem[\protect\citeauthoryear{Moore et al.}{1999}]{Moore} Moore B. et al. 1999, Aph.J., \textbf{524}, L19.

\bibitem[\protect\citeauthoryear{Navarro et al.}{1996}]{NFW} Navarro J., Frenk C., White S.  1996, ApJ, \textbf{462}, 563.

\bibitem[\protect\citeauthoryear{Neindorf}{2003}]{Neindorf} Neindorf B.,  2003, A\&A, \textbf{404}, 83.

\bibitem[\protect\citeauthoryear{Oguri}{2005}]{O} Oguri M. 2005, MNRAS, \textbf{361(1)}, L38.

\bibitem[\protect\citeauthoryear{Oort}{1932}]{OO} Oort J.H., 1932, Bull. Astron. Inst. Netherlands, \textbf{6}, 249.

\bibitem[\protect\citeauthoryear{\"Opik}{1932}]{Opik} \"Opik E. 1915, Bull. de la Soc. Astr. de Russie \textbf{21}, 150.

\bibitem[\protect\citeauthoryear{Paczy\'nski}{1986a}]{P} Paczy\'nski B. 1986, Aph.J. \textbf{301}, 503.
\bibitem[\protect\citeauthoryear{Paczy\'nski}{1986b}]{Paczyn} Paczy\'nski B. 1986, Aph.J. \textbf{304}, 1.

\bibitem[\protect\citeauthoryear{Paczy\'nski \& Wambsganss}{1989}]{PaczWambs}    Paczy\'nski B., Wambsganss J. 1989,  ApJ \textbf{337}, 581.

\bibitem[\protect\citeauthoryear{Paduroiu et al.}{2015}]{PRP} Paduroiu S., Revaz Y., Pfenniger D. 2015,  ArXiv 1506.03789.

\bibitem[\protect\citeauthoryear{Riess et al.}{2014}]{RCA} Riess A.G., Casertano S., Anderson J., MacKenty J., Filippenko A. 2014,  Aph.J., \textbf{785(2)}, 161.

\bibitem[\protect\citeauthoryear{Rocha et al.}{2012}]{R} Rocha M.E. et al. 2012, MNRAS, \textbf{430(1)}, 81.

\bibitem[\protect\citeauthoryear{Salpeter}{1955}]{salpeter} Salpeter E., 1955, ApJ, \textbf{121}, 161.

\bibitem[\protect\citeauthoryear{Schechter \& Wambsganss}{2002}]{ScW} Schechter P., Wambssganss J. 2002, AJ, \textbf{580}, 685.

\bibitem[\protect\citeauthoryear{Schneider et al.}{1992}]{SchnEhlFal} Schneider P.,  Ehlers J., and  Falko E. E., Gravitational Lenses, Springer, New York, NY, USA, 1992.

\bibitem[\protect\citeauthoryear{Schneider et al.}{2010}]{SKM1} Schneider A., Krauss L., Moore B. 2010, Phys.Rev.D, \textbf{82}, 063525.

\bibitem[\protect\citeauthoryear{Schneider et al}{2011}]{SKM2} Schneider A., Krauss L., Moore B. 2011, ArXiv 1105.4106.

\bibitem[\protect\citeauthoryear{Schneider et al.}{2012}]{SSMM} Schneider A., Smith R.E., Macci A., Moore B. 2012, MNRAS, \textbf{424}, 684.

\bibitem[\protect\citeauthoryear{Schneider \& Weiss}{1987}]{SW} Schneider P., Weiss A., 1987, MPA Rep., \textbf{311}, 46.

\bibitem[\protect\citeauthoryear{Schmidt \& Wambsganss}{2010}]{Schmidt2010} Schmidt R. W., Wambsganss J., 2010, GRG \textbf{42}, 2127.

\bibitem[\protect\citeauthoryear{Schaw et al.}{2007}]{SWOB} Shaw L. D., Weller J., Ostriker J.P., Bode P. 2007, Aph.J., \textbf{659}, 1082.

\bibitem[\protect\citeauthoryear{Seitz et al.}{1994}]{Seitz1994} Seitz C., Schneider P. 1994, A\&A \textbf{288}, 1.

\bibitem[\protect\citeauthoryear{Shin et al.}{2007}]{Shin}  Shin I.-G.; Han C.; Choi J.-Y. et al. 2012, Aph.J.,   \textbf{755}, id. 91, 10 pp.

\bibitem[\protect\citeauthoryear{Shull}{2014}]{Shull} Shull J.M. 2014, Aph.J., \textbf{784}, Is. 2, 10.

\bibitem[\protect\citeauthoryear{Sliusar et al.}{2015}]{KFNT} Sliusar V.N.,  Zhdanov V.I.,  Alexandrov A.N., Fedorova E.V. 2015, Kinemat. Phys. Celest. Bodies \textbf{31}, Is. 2, 47.

\bibitem[\protect\citeauthoryear{Springel et al.}{2005}]{SXXXX} Springel V. et al. 2005, Nature, \textbf{435}, 629.

\bibitem[\protect\citeauthoryear{Springel et al.}{2008}]{Springel}  Springel V., Wang J., Vogelsberger M. et al. 2008, MNRAS, \textbf{391}, 1685.

\bibitem[\protect\citeauthoryear{Stadel et al.}{2009}]{Stadel} Stadel J.,  Potter D., Moore B., et al. 2009, MNRAS, \textbf{398}, L21.

\bibitem[\protect\citeauthoryear{Swaters et al.}{2011}]{Swaters} Swaters R.A., Sancisi R., van Albada T.S., van der Hulst J.M. 2011, Aph.J., \textbf{729}, DOI:10.1088/0004-637X/729/2/11.

\bibitem[\protect\citeauthoryear{Tisserand et al.}{2007}]{EROS}  Tisserand P.,  Le Guillou L.,  Afonso C., et al. 2007, A\& A, \textbf{ 469}, 387.

\bibitem[\protect\citeauthoryear{Udalski et al.}{1993}]{Udalski} Udalski A., Szyma\'nski M., Ka\l u\.zny J., et al. 1993, Acta Astron., \textbf{43}, 289.

\bibitem[\protect\citeauthoryear{Wambsganss et al.}{1990}]{WambPaKa}  Wambsganss, J., Paczy\'nski, B., Katz, N. 1990, Aph.J \textbf{352}, 407.

\bibitem[\protect\citeauthoryear{Wambsganss et al.}{1992}]{Wambs}     Wambsganss J., Witt H.J., Schneider P. 1992, A\&A, \textbf{258}, 591.

\bibitem[\protect\citeauthoryear{Wyrzykowski et al.}{2009}]{OGLE}  Wyrzykowski \L.,  Koz\l owski S., Skowron J., et al., 2009, MNRAS, \textbf{397}, 1228.

\bibitem[\protect\citeauthoryear{Vogelsberger et al.}{2012}]{Vog} Vogelsberger M. et al. 2012, MNRAS, \textbf{423}, 3740.

\bibitem[\protect\citeauthoryear{Zackrisson \& Riehm}{2010}]{ZR}  Zackrisson E., Riehm T. 2010, Advances in Astronomy, id. 478910.

\bibitem[\protect\citeauthoryear{Zakharov}{2010}]{Z} Zakharov A. 2010, Gen.Rel.Grav., \textbf{42}, 2301.

\bibitem[\protect\citeauthoryear{Zakharov \& Sazhin}{1999}]{ZS} Zakharov A.F., Sazhin M.V. 1999, A\&AT, \textbf{18}, 27.

\bibitem[\protect\citeauthoryear{Zhdanov et al.}{2012}]{ISRN_Zh} Zhdanov V., Alexandrov A., Fedorova E., Sliusar V. 2012, ISRN A\&A, \textbf{2012}, ID 906951.

\bibitem[\protect\citeauthoryear{Zhang}{2011}]{Zhang} Zhang D. 2011, MNRAS, \textbf{418(3)}, 1850.

\bibitem[\protect\citeauthoryear{Zwicky}{1933}]{Zwicky} Zwicky F. 1933, Helvetica Physica Acta, \textbf{6}, 110.


\end{thebibliography}
\end{document}